\documentclass[12pt]{iopart}
\usepackage{amssymb}
\usepackage{graphicx}
\usepackage{color}

\begin{document}
\title[Theory of cavity-assisted microwave cooling of polar molecules.]
{Theory of cavity-assisted microwave cooling of polar molecules. }

\author{Margareta Wallquist$^1$, Peter Rabl$^2$, Mikhail D. Lukin$^2$ and Peter Zoller$^1$}
\address{$^1$Institute for Theoretical Physics, University of Innsbruck, and \\
Institute for Quantum Optics and Quantum Information of the Austrian Academy of Sciences, 6020 Innsbruck, Austria.}
\address{$^2$Institute for Theoretical Atomic, Molecular and Optical Physics,
Cambridge, MA 02138, USA, and \\ Physics Department, Harvard University, Cambridge, MA 02138, USA.}
\ead{margareta.wallquist@uibk.ac.at}
\begin{abstract}
We analyze cavity-assisted cooling schemes for polar molecules in the microwave domain, where molecules are excited on a rotational transition and energy is dissipated via strong interactions with a lossy stripline cavity, as recently proposed by A. Andr\'e {\it et al.}, Nature Physics {\bf 2}, 636 (2006). We identify the dominant cooling and heating mechanisms in this setup and study cooling rates and final temperatures in various parameter regimes. In particular we analyze the effects of a finite environment temperature on the cooling efficiency, and find minimal temperature and optimized cooling rate in the strong drive regime. Further we discuss the trade-off between efficiency of cavity cooling and robustness with respect to ubiquitous imperfections in a realistic experimental setup, such as anharmonicity of the trapping potential.
\end{abstract}
\maketitle

\section{Introduction}
Cold molecules are nowadays very actively explored~\cite{SpecialIssue04}. Of particular interest are cold heteronuclear molecules prepared in their electronic and vibrational ground state, which in view of their electric dipole moments can provide a strong dipole-dipole interaction, and a strong coupling of rotational excitations  to microwave fields and cavities. The field of potential applications of cold samples of polar molecules ranges over ultracold  chemistry~\cite{KremsRPC2005} and precision measurements \cite{HudsonPRL2002,DeMillePRL2008}, quantum computation~\cite{DeMillePRL2002,YelinPRA2006} and the simulation of exotic condensed matter models~\cite{WangPRL2006,MicheliNatureP2006,BuechlerNatureP2007}. A necessary prerequisite for these developments is the ability to cool molecular ensembles and single trapped molecules to very low temperatures.

Several techniques for cooling, and preparation of ultracold molecular ensembles have been investigated~\cite{DoyleReview2004}  both experimentally and theoretically, including cavity-assisted laser cooling~\cite{VuleticPRA2001,MorigiPRL2007,LevPREPRINT}, buffer gas cooling, and Stark deceleration of polar molecules including trapping in magnetic and electric traps~\cite{MeijerReview2006,CampbellPRL2007,TarbuttPRL2004,Rieger05,MeekPREPRINT}, evaporative~\cite{AvdeenkovPRA2006} and sympathetic cooling schemes~\cite{LaraPRL2006},  and preparation of ground state molecules by photoassociation of cold atoms~\cite{Mancini04,Sage05,Wang04,KleinertPRL2007}. Recently a technique for  trapping and cooling of polar molecules on a chip has been suggested, which relies on  strong coupling of rotational states of the molecule to a superconducting stripline cavity~\cite{AndreNaturePh2006}. When trapped at a distance $d \sim 1 \,\mu$m above the cavity electrodes, the vacuum Rabi frequency $g$ associated with the exchange of a rotational excitation and a single microwave photon is in the order of $100$ kHz.
In combination with a comparable photon loss rate $2\kappa \sim g$ this coupling serves as an efficient dissipation channel for (the in free space essentially stable) rotational states, which is exploited for cooling the motion of a trapped molecule.  Cooling forces in this setup arise from the strong gradients of microwave fields in the near-field regime, in analogy to cavity-assisted laser cooling techniques discussed for atoms in the optical regime~\cite{VuleticPRA2001,CiracPRA1995,HorakPRL1997,ZippilliPRL2005,ZippilliPRA2005,MaunzNature2004,BoozerPRL2006}. Microwave cooling does not rely on specific electronic or collisional properties~\cite{KremsRPC2005,AvdeenkovPRA2006,LaraPRL2006,TscherbulPRL2006,Krems2004,TicknorPRA2005} of the molecules and therefore should be applicable for a large set of di- and also multiatomic polar molecules. As molecules are prepared in the motional ground state of a chip-based trap, this cooling technique is in particular interesting for applications in the context of on-chip (hybrid) quantum information processing~\cite{AndreNaturePh2006,RablPRL2006}.

In this work we will present a detailed theoretical study of cavity assisted cooling  of a single trapped polar molecule in the microwave domain. The goal of our theoretical analysis is to identify the physical mechanisms behind different cooling schemes, and to discuss the cooling rates and final temperatures in various parameter regimes including weak and strong driving fields and a finite temperature of the environment. The main conclusions of this analysis are as follows. Depending on the detuning between the relevant rotational transition frequency and the microwave fields, cavity cooling is dominated either by cavity assisted sideband cooling (CASC), where cooling forces arise from the gradient of a classical driving field, or ``$\nabla g$" cooling where forces are due to gradients in the coupling between the molecule and a single microwave photon.
Zero environmental temperature in principle allows ground state cooling in the resolved sideband limit for the respective processes.
At finite temperature $T$, thermal cavity photons at frequency $\omega_{\rm c}$ introduce an additional heating mechanism, and the minimal temperature $T_{\rm f}$ is given by $T_{\rm f} = (\nu/\omega_{\rm c}) T \ll T$ with $\nu$ the trap frequency. We find that this limit can be reached in the strong driving regime, where also cooling rates are optimized. Our analysis of cavity cooling in the presence of imperfections such as anharmonicity of the trapping potential shows that by switching between the CASC and the $\nabla g$ cooling regime, cooling efficiency is traded for an increased robustness of the cooling scheme. This may in particular be important for ongoing experiments.

The paper is structured as follows. In section \ref{sec:Overview} we introduce the model for a trapped polar molecule coupled to a stripline cavity, and present a brief overview of the basic physical ideas and mechanisms underlying the cavity assisted microwave cooling. In section \ref{sec:Theory} we proceed with a technical derivation of the cooling and heating rates in the bad cavity limit. Based on these results we present in section \ref{sec:Discussion} a more detailed discussion of cooling rates and final temperatures as a function of different system parameters. Finally, in section \ref{sec:Summary} we summarize the main results and conclusions of this work.
%

\section{ Overview on cavity-assisted microwave cooling of a trapped polar molecule}\label{sec:Overview}
The goal of this work is to present a theory for ground-state cooling of trapped polar molecules, in analogy to the ideas and theoretical tools employed for cooling of trapped ions in the optical domain~\cite{StenholmRMP1986,LeibfriedRMP2003}.
In this section we start with a brief review of the system involving a polar molecule coupled to a stripline cavity, as proposed in \cite{AndreNaturePh2006}. Further, the general relation between cooling/heating rates and the spectrum of fluctuations of the dipole forces which act on the molecule allows us to identify different cooling mechanisms in this setup. A qualitative discussion of the individual mechanisms is presented together with analytical results for a weakly driven molecule. The goal of this overview is to provide a physical understanding of the main cooling and heating processes which serves as a guideline through the more technical derivation of the force spectrum in section \ref{sec:Theory} and the extended discussion in section \ref{sec:Discussion}.

\subsection{Cavity QED with a single trapped polar molecule coupled to a microwave cavity}\label{sec:Model}

\begin{figure}
\begin{centering}
\includegraphics[width=0.85\textwidth]{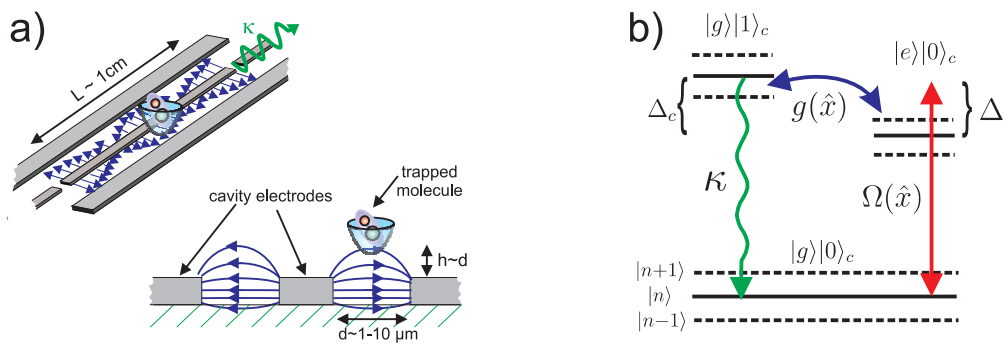}
\caption{a) Schematic plot of the setup considered in this paper for cavity-assisted cooling of polar molecules. A more realistic trap design can be found elsewhere \cite{AndreNaturePh2006}. b) Level diagram for the cavity-molecule system restricted to the lowest states $|g\rangle|0\rangle_{\rm c}$, $|e\rangle|0\rangle_{\rm c}$ and $|g\rangle|1\rangle_{\rm c}$. Dashed lines indicate different quantized motional levels. See text for more details.   } \label{fig:Setup}
\end{centering}
\end{figure}

We consider the setup shown in figure \ref{fig:Setup}, where a single polar molecule is trapped close to a superconducting stripline cavity.
The system Hamiltonian $H_{\rm S}$,
\begin{equation}
H_{\rm S}= H_{\rm c} +H_{\rm m} +  H_{\rm int}\,,
\label{eq:HsGen}
\end{equation}
is given in terms of the Hamiltonian $H_{\rm c}$ for the microwave cavity, the Hamiltonian $H_{\rm m}$ describing the internal degrees of freedom of the molecule including a classical microwave driving field, as well as its motion (external degree of freedom) in the trap. Finally $H_{\rm int}$ describes the dipole interaction  between rotational states of the molecule and the electric field of the cavity. Apart from the coherent evolution governed by $H_{\rm S}$, coupling of microwave photons to the electromagnetic environment introduces a dissipative decay channel for cavity photons. In section \ref{sec:Theory} we incorporate dissipation in terms of a master equation description of our system. In the following we summarize the properties of $H_{\rm S}$ which are relevant for the discussion in this paper while more details can be found in \cite{AndreNaturePh2006,RablPRA2007}.

With initial temperatures of the molecule in the order of 1-100 mK, electronic and vibrational degrees of freedom are frozen out and the molecule is well described by its rotational degrees of freedom only.  The anharmonicity of the rotor spectrum, $E_{\rm rot}(N) = BN(N+1)$, with $B$ the rotational constant and $N$ the angular momentum quantum number, allows us to identify two specific rotational states $|g\rangle$ and $|e\rangle$ with a transition frequency $\omega_{eg}=(E_e-E_g)/\hbar$ typically in the tens of GHz regime. With spontaneous emission rates of low lying rotational states in the mHz regime for zero temperature, the idea is to create an effective decay channel for the excited state $|e\rangle$ via a microwave stripline cavity. When the frequency of the relevant cavity mode $\omega_{\rm c}$ is near  resonance with $\omega_{eg}$, the dynamics of the molecule is restricted to the two states $\{ |g\rangle, |e\rangle\}$ and the molecule is well approximated by a two-level system (TLS).

The microwave cavity sketched in figure \ref{fig:Setup} is a (quasi-) one-dimensional superconducting stripline with length $L \sim 1$ cm, and transverse dimension of order $d\sim 1\,\mu$m. Allowed wavevectors $k_n=n\pi/L$ correspond to a mode spacing in the GHz regime.
Restricted to a single resonator mode,
the cavity Hamiltonian is $H_{\rm c}=\hbar \omega_{\rm c} c^\dag c $ with $c$ ($c^\dag$) the annihilation (creation) operators for photons of frequency $\omega_{\rm c}$. The electric field generated by the cavity is given by the operator $\hat E_{\rm c} =\mathcal{E}_0(\vec r) (c+c^\dag)$ with $\mathcal{E}_0(\vec r)$ the field strength associated with a single microwave photon. While $\mathcal{E}_0(\vec r)$ varies on the scale of  $L$ along the cavity axis, the spacial dependence of $\mathcal{E}_0(\vec r)$ in the transverse direction or at the cavity end-caps is determined by the typical electrode spacing $d$. For strong cooling forces we are in particular interested in those strong gradients in the near field regime. The microwave photons dissipate through the coupling of the stripline to its surrounding leads. At frequencies of $\omega_{\rm c}\sim 10$ GHz, quality factors in the range of $Q \sim 10^3 - 10^6$~\cite{DayNature2003,WallraffNature2004,FrunzioIEEE2005} translate into a photon loss rate $2\kappa$ in the range of MHz to kHz.

Molecules interact with microwave photons via the dipole interaction $H_{\rm int} = - \hat{\mu} \cdot \hat{E}_{\rm c}$, where $\hat \mu$  is the electric dipole operator of the molecule. The rotational states $|g\rangle$ and $|e\rangle$ are coupled to the quantized electric field of the cavity. Disregarding, for the moment, the motional degrees of freedom we can  describe the combined cavity-molecule system, including a classical microwave field of frequency $\omega_{\rm m}$ and Rabi-frequency $\Omega$ which drives transitions between $|g\rangle$ and $|e\rangle$,
by a Jaynes-Cummings type model in a frame rotating with the microwave frequency $\omega_{\rm m}$,
\begin{equation}
H_{\rm JC} = \hbar \left( \Delta_{\rm c} - \Delta \right) c^\dag c - {\hbar \Delta \over 2}\,\sigma_z +   \hbar g \left( \sigma_+ c + \sigma_- c^\dagger \right)+ {\hbar \Omega \over 2} \,\sigma_x ,
\label{Eq:JC}
\end{equation}
where $\Delta_{\rm c}=\omega_{\rm c}-\omega_{eg}$ and $\Delta=\omega_{\rm m}-\omega_{eg}$ are the detunings of the cavity field and the classical driving field from the  rotational transition frequency, respectively, and we have adopted the usual Pauli operator notation for the TLS. Under resonance conditions $\omega_{c}\approx \omega_{eg}$ equation
(\ref{Eq:JC}) describes coherent oscillations between an excitation in the TLS, $|e\rangle |n\rangle_{\rm c}$, and an additional photon in the cavity, $|g\rangle |n+1\rangle_{\rm c}$, on a timescale $(g\sqrt{n+1})^{-1}$ ~\cite{HoodScience2000,RaimondRMP2001}.  For a molecule trapped at a distance $h\sim d\sim 1-0.1\,\mu$m above the cavity and a molecular dipole moment of the order of several Debye the vacuum Rabi frequency $g$ is of the order of $2\pi\times 50-500$ kHz
\cite{AndreNaturePh2006}.
In combination with this coherent energy transfer between molecule and cavity, the photon decay opens an effective decay channel for the rotationally excited state $|e\rangle$. For  $\kappa > g $ the corresponding characteristic decay rate is given by
 \begin{equation}\label{eq:gamma}
\gamma =  {2g^2 \kappa \over \kappa^2 + \Delta_{\rm c}^2}\,.
\end{equation}
Although the validity of this expression is limited to $\gamma \lesssim g$, we find a significant increase of the rotational decay rate from $\Gamma_{eg}\sim$ mHz in free space to a cavity enhanced decay of $\gamma\sim 10-100$ kHz. This also implies that in principle cooling rates of the same order can be achieved in this system.

Trapping of a molecule close to the surface of a chip requires a sufficiently strong trapping potential; electrostatic traps with frequencies of the order of $\nu/2\pi\sim 1$ MHz are experimentally feasible \cite{AndreNaturePh2006} \footnote{In principle our derivation applies equally well to magnetic traps, as used in current atom chip experiments \cite{AtomChips}, in the case of a molecule with an unpaired electron spin. Note however that the trap fields must be compatible with the superconducting setup.}.
While in general the trapping potentials will be different for the two states $|g\rangle$ and $|e\rangle$, there exist so-called `sweet spots' for the electric bias fields, where this difference vanishes \cite{AndreNaturePh2006}. Let us consider a 1D model in which case the external degrees of freedom are described by a harmonic oscillator Hamiltonian $H_{\rm E}=\hbar \nu a^\dag a$, with $a$ ($a^\dag$) the usual annihilation (creation) operators. Coupling between internal and external degrees of freedom arises from the spatial dependence of the Rabi frequencies $g\rightarrow g(\hat x)$ and $\Omega\rightarrow \Omega(\hat x)$, as compared to the case of laser cooling which we already remarked in the introduction. Here $\hat x$ is the position operator of the molecule with respect to the center of the trap.
Including motional degrees of freedom into equation (\ref{Eq:JC}), we end up with the explicit form for the system Hamiltonian $H_{\rm S}$ (\ref{eq:HsGen}),
\begin{eqnarray}
H_{\rm c} &=& \hbar (\Delta_{\rm c}-\Delta)c^\dag c , \nonumber \\
H_{\rm m} &=& \hbar \nu  a^\dag a - {\hbar \Delta \over 2}\, \sigma_z + {\hbar\Omega(\hat x) \over 2} \,\sigma_x\,, \nonumber \\
H_{\rm int} &=& \hbar g(\hat{x}) \left( \sigma_+ c + \sigma_- c^\dagger \right) .
\label{Eq:Hs}
\end{eqnarray}
The bare energy levels of $H_{\rm S}$ and the different couplings between them are sketched in figure~\ref{fig:Setup} b). We emphasize at this point that the validity of $H_{\rm m}$ given in equation (\ref{Eq:Hs}) is based on two crucial assumptions: i) the trapping potential is harmonic and ii) the trapping potential is state independent. Assumption i) is in general fulfilled near the trap center, i.e., in the final stage of the cooling process and as mentioned above, assumption ii) is fulfilled for certain conditions on the trapping fields. As neither of those assumptions will be strictly satisfied in a realistic experiment, we will investigate small deviations from these conditions in section~\ref{sec:Discussion}. Note that as long as condition i) is approximately fulfilled we can treat the motion of the molecule along the $x$, $y$ and $z$ directions independently. This justifies the use of the one dimensional model for the motion of the molecule.

While for all cooling schemes in the optical domain the temperature of the electromagnetic environment can safely be set to zero, this assumption is in general not valid in the microwave regime. In a cryogenic environment with a temperature $T \sim 0.02 -  4\,K$ the thermal occupation number of the cavity mode, $N=1/ [\exp(\hbar\omega_{\rm c}/k_{\rm B}T)-1]$, is typically of the order of $N\sim  0 - 10$. In addition to the vacuum contribution in equation (\ref{eq:gamma}), a finite $N$ results in cavity-assisted decay of rate $\gamma(N+1)$ and a cavity-assisted excitation rate $\gamma N$. As we discuss in more detail below, thermally activated processes are one of the limiting factors in achieving cold temperature.

\subsection{Cavity cooling of polar molecules in the Lamb-Dicke regime}

When trapped close to the surface of the chip, cooling forces on the molecule arise from variations of the microwave fields on the typical length scale $d\sim 1\,\mu$m, which is only determined by the electrode geometry. For a harmonic trapping potential we compare this scale with the harmonic oscillator length $x_0=\sqrt{\hbar/2m\nu}$ and introduce the dimensionless parameters $\eta\equiv x_0 \nabla \Omega /\Omega |_{x=0}$ and $\eta_{\rm c}\equiv x_0 \nabla g /g |_{x=0}$ in analogy to the Lamb-Dicke parameter in optical cooling schemes; e.g. in the case of trapped ions, $\eta = 2\pi x_0/\lambda$ with $\lambda $ the laser wavelength \cite{StenholmRMP1986}. As the linear extension of the molecule trap is approximately limited by $d$ we find that cooling in this setup naturally occurs in the so-called Lamb-Dicke limit $\eta,\eta_{\rm c}\approx x_0/d \ll 1$, where the spatial extension of the molecule wavepacket is small compared to variations of the microwave fields.  While typically $\eta$ and $\eta_{\rm c}$ will be equal or at least of the same order, in the following discussion we treat them as two independent parameters.

In the Lamb-Dicke limit we can expand the Hamiltonian (\ref{Eq:Hs}) to first order in $\hat x$, i.e. $H_{\rm S}\simeq H_0+H_1$. Here $H_0$ is the Hamiltonian describing the free evolution of the internal and motional degrees of freedom of the molecule. The coupling between the internal and motional degrees of freedom is of the form
\begin{equation}
H_1=\hat x\hat F, \qquad  \hat F =  \frac{\hbar \nabla \Omega}{2}{\sigma_x}+ \hbar\nabla g(\sigma_+ c + \sigma_-c^\dag )\,,
\end{equation}
such that  $\langle \hat F\rangle$ is the mean dipole force exerted on the molecule.
The small parameters $\eta$ and $\eta_{\rm c}$ allow us to treat $H_1$ as a weak perturbation for the decoupled dynamics of the system. For $|H_1|\ll \Gamma_{\rm I}$, with $\Gamma_{\rm I}={\rm  min}\{\gamma,\kappa\}$ the characteristic relaxation rate of the cavity-molecule system, we can adiabatically eliminate the internal degrees of freedom (see section \ref{sec:Theory}) and derive an effective equation of motion for the reduced external density operator $\mu(t)$.
The resulting equation for the mean occupation number  $\langle n\rangle(t)={\rm Tr}\{ a^\dag a  \mu (t)\}$ (see also \cite{ZippilliPRA2005,StenholmRMP1986,CiracPRA1992}) is of the form
\begin{equation}
\langle \dot n\rangle = - W \left[\langle n\rangle-\langle n\rangle_0\right].
\label{Eq:dotn}
\end{equation}
For a positive damping rate $W>0$, equation (\ref{Eq:dotn}) describes relaxation of the mean occupation number to a final steady state value $\langle n\rangle_0$. The relevant quantities characterizing the cooling dynamics, $W$ and $\langle n\rangle_0$, are given by
\begin{equation}
W=A_--A_+,\qquad \langle n\rangle_0 = A_+/(A_--A_+)\,,
\end{equation}
where the cooling and heating rates $A_-=S(\nu)$ and $A_+=S(-\nu)$ are determined by the spectrum of the force operator,
\begin{equation}
S(\omega)=\frac{2 x_0^2}{\hbar^2} \,{\rm Re}\int_0^\infty d\tau \, \langle \hat F(\tau) \hat F(0)\rangle_0 \, e^{+i\omega t}\,.
\label{Eq:Sgeneral}
\end{equation}
Equations (\ref{Eq:dotn})-(\ref{Eq:Sgeneral}) provide the general relation between the cooling dynamics of the external degrees of freedom and the fluctuation spectrum of the internal operator $\hat F$.
Efficient cooling schemes correspond to the case where both the difference $S(\nu)-S(-\nu)$ is maximized (large cooling rate) and the ratio $S(-\nu)/S(\nu)$ is minimized (low final temperatures). Throughout this paper we make extensive use of this relation to discuss different cooling schemes in terms of the properties of $S(\omega)$.

\subsection{Cavity cooling of polar molecules in the bad cavity regime}

 While the results for cooling in the Lamb-Dicke limit summarized in equations (\ref{Eq:dotn})-(\ref{Eq:Sgeneral}) are valid for arbitrary cavity-molecule interactions, we concentrate in the remainder of this paper on the bad cavity limit (BCL), $g\ll\kappa$, which we expect to be the most relevant regime in experiments. For additional discussions on cavity cooling of atoms in the good cavity limit the reader is referred to \cite{ZippilliPRA2005}.
\begin{figure}
\begin{centering}
\includegraphics[width=0.9\textwidth]{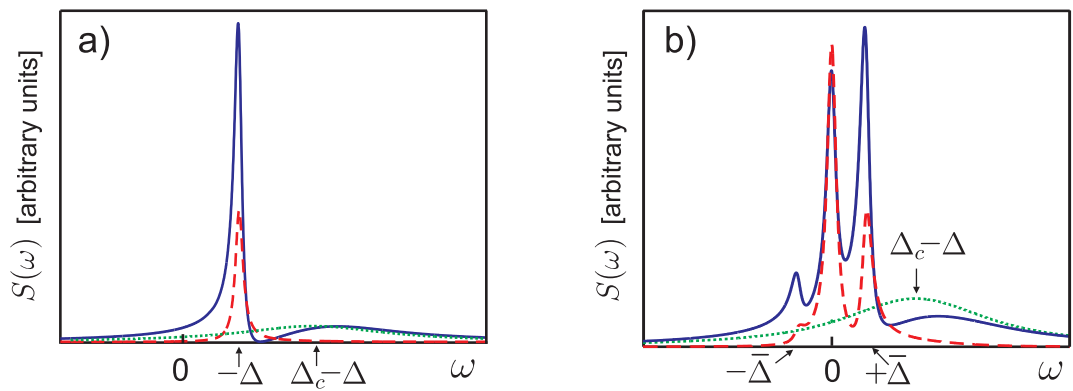}
\caption{Example for the force spectrum $S(\omega)$ for zero temperature ($N=0$) and for a) $\Omega \ll |\Delta|$ and  b) $\Omega \sim |\Delta|$. In both figures the contribution of $S_\Omega(\omega)$ and $S_g(\omega)$ are indicated by the dashed and dotted lines, respectively.} \label{fig:SpecExample}
\end{centering}
\end{figure}
The BCL allows us to eliminate the cavity degrees of freedom and to study and interpret cavity-cooling in terms of the resulting effective dynamics for the molecule. However, based on the underlying coherent interaction and the finite bandwidth of the cavity mode, the effective dynamics of the molecule differs significantly from the usual case of a two level atom coupled to the free radiation field. The adiabatic elimination of the cavity mode, which is discussed in more detail in section \ref{sec:Theory}, results in a decomposition of the general form of $S(\omega)$ given in equation (\ref{Eq:Sgeneral}) into three contributions,

\begin{equation}
S(\omega)= S_{\Omega}(\omega)  +S_{g} (\omega)+ S_{I}(\omega)\,.
\label{Eq:S_contr}
\end{equation}
The labels on the first two terms refer to their origin; the spectrum $S_\Omega$ is associated with the force from the drive field gradient $\nabla\Omega$ and exhibits sharp resonances of width $\sim\gamma$, whereas $S_g$ is associated with the force due to the cavity-TLS coupling gradient, $\nabla g$, and exhibits a broader resonance of width $2\kappa$. As explained in more detail in the following discussion, we can with each of these forces associate a different cooling mechanism in our system. The third contribution to the spectrum, $S_{I}$, describes destructive and constructive interference between the different cooling processes. For illustration, the spectrum $S(\omega)$ and its decomposition according to equation (\ref{Eq:S_contr}) are shown in figure \ref{fig:SpecExample} for the case of zero temperature and a) for a weakly ($\Omega \ll |\Delta|$) and b) for a strongly ($\Omega \sim |\Delta|$) driven molecule.

The results presented in this overview section apply to a weakly driven molecule, and generalizations to arbitrary $\Omega$ are derived in section \ref{sec:Discussion}.

\subsubsection{Cavity-assisted sideband/Doppler cooling.}
\begin{figure}
\begin{centering}
\includegraphics[width=0.9\textwidth]{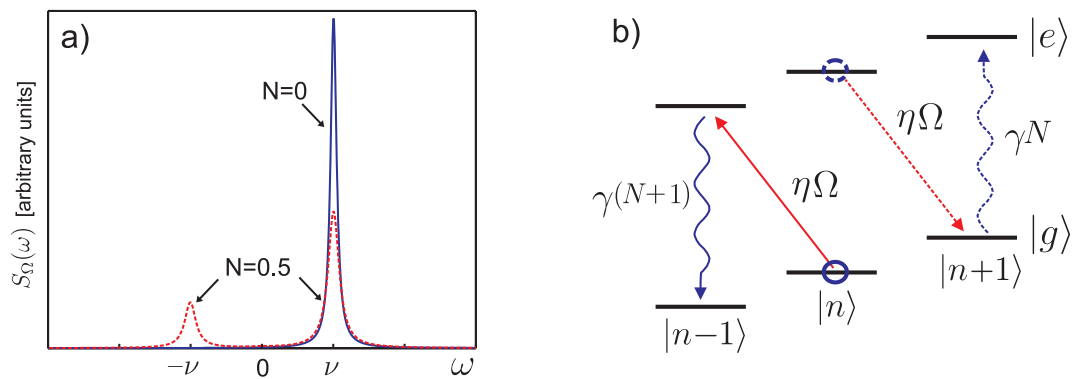}
\caption{Cavity-assisted sideband cooling. a) Spectrum $S_\Omega(\omega)$ in the weak driving regime $\Omega \ll |\Delta|$ for $N=0$ (solid line) and $N=0.5$ (dashed line). The parameters chosen for this plot are (all in units of $\nu$):  $\Omega=0.1$, $\Delta=-1$, $\Delta_{\rm c}=0$, $\kappa=5$, $g=0.5$.  b) Resonant processes contributing in the sideband resolved regime. Dashed lines indicate processes at finite environment temperature.} \label{fig:SepcExample}
\end{centering}
\end{figure}

We first consider the limit $\eta_{\rm c}\rightarrow 0$ where the cavity field exerts no force on the molecule but still provides an effective
decay channel for the excited state $|e\rangle$. In this limit the force spectrum reduces to the first term in equation (\ref{Eq:S_contr}), $S(\omega)\equiv S_\Omega(\omega)$.
For zero temperature and $\Omega \ll \gamma$  the spectrum exhibits a single Lorentzian peak of width $\gamma$ centered at a frequency $\omega=-\Delta$ (see figure \ref{fig:SepcExample}). We identify $S_\Omega(\omega)$ with the excitation spectrum of a decaying TLS and recover a situation which is familiar from laser cooling of trapped ions \cite{StenholmRMP1986,LeibfriedRMP2003}. In the so-called resolved sideband limit, i.e. when the trap frequency is large compared to the effective decay rate $\gamma$, we can choose the detuning $\Delta=-\nu$ such that $S(\nu)\gg S(-\nu)$. Under those conditions sideband cooling leads to final occupation numbers close to the quantum ground state, $\langle n\rangle_0=(\gamma/4\nu)^2$.

For non-zero temperature the cavity acts as a finite temperature reservoir for the molecule, which has a corresponding equilibrium excited state population $\rho_{ee}^0=N/(2N+1)$. In the spectrum another resonance appears at $\omega=-\nu$ with a height $\sim \rho_{ee}^0$. As seen from figure \ref{fig:SepcExample} b) this resonance corresponds to transitions from the excited state to the ground state which for $\Delta=-\nu$ preferably increase the motional quantum number and lead to heating. In equilibrium heating and cooling rates balance each other, meaning that the steady state occupation number $\langle n\rangle_0$ in the trap is equal to the thermal cavity occupation number $N$. This equilibrium condition translates into a final temperature $T_{\rm f}=(\nu/\omega_{\rm c})T \ll T$.
In section \ref{sec:Discussion} we show that for a weak driving field $\Omega \ll \gamma_N = (2N+1)\gamma$ and for $\gamma_N \ll \nu$ the correct interpolation between the zero temperature limit $\langle n \rangle_0 = (\gamma/4\nu)^2$ and the finite temperature result $\langle n \rangle_0 \simeq N$ of sideband cooling is given by
\begin{equation}
\langle n \rangle_0 = N + (2N+1) \left({\gamma_N \over 4 \nu}\right)^2.
\label{eq:NfSB}
\end{equation}
While in the sideband resolved regime cooling is most efficient we may instead consider the so-called Doppler limit $\gamma_N \gg \nu$. In this regime cooling rates are optimized by choosing the detuning $\Delta = -\gamma_N/2$ and the minimum final occupation number is,
\begin{equation}
\langle n \rangle_0 = (2N+1) {\gamma_N \over 4 \nu}.
\label{eq:NfD}
\end{equation}
 In contrast to the sideband resolved regime, we find that when the linewidth $\gamma_N$ exceeds the trapping frequency $\nu$ the scaling of the final occupation number with temperature is less favorable $(\sim N^2)$ and cooling becomes inefficient for large $N$.

Equations (\ref{eq:NfSB}) and (\ref{eq:NfD}) have been derived for a weakly driven molecule and generalize the usual limits for sideband and Doppler cooling of trapped ions to the case of a finite temperature reservoir. In section~\ref{sec:Discussion} we study the limit of a strongly driven molecule (see figure~\ref{fig:SpecExample} b)), where we show that
by increasing $\Omega$ we benefit not only from a higher damping rate $W$, but cavity assisted sideband cooling also gets more robust with respect to imperfections in the trapping potential.

\subsubsection{$\nabla g$-cooling. }
We now consider the opposite limit $\eta\rightarrow 0$ where cooling forces arise from gradients of the cavity-molecule coupling $g(\hat x)$ only. In addition, to avoid interference effects discussed below, for the moment we assume $g\rightarrow 0$ such that the spectrum of force fluctuations reduces to the second term in equation (\ref{Eq:S_contr}),   $S(\omega)\equiv S_g(\omega)$.

\begin{figure}
\begin{centering}
\includegraphics[width=0.9\textwidth]{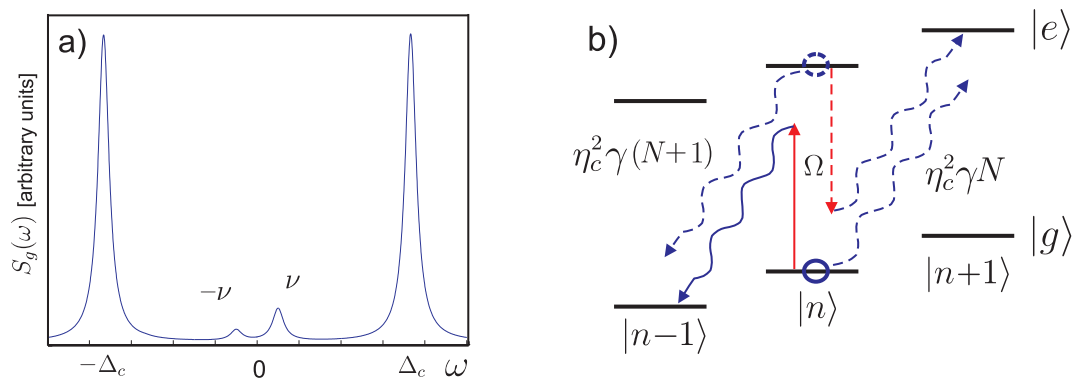}
\caption{Spectrum $S_g(\omega)$ in arbitrary units for the set of parameters (all in units of $\nu$):
$g=0.01$, $\kappa=0.3$, $\Omega=2$, $\Delta=6$, $\Delta_{\rm c}=7$, $N=0.5$ b) processes which contribute  to the individual resonances in the sideband resolved regime.   } \label{fig:SepcExampleGG}
\end{centering}
\end{figure}

For simplicity we first look at the case of zero temperature $N=0$. As shown in figure \ref{fig:SpecExample}, $S_g(\omega)$ has a single resonance at $\omega = \Delta_{\rm c}-\Delta$ with a width  equal to the photon loss rate $2\kappa$. The deviation from the naively expected value $\omega=\Delta_{\rm c}$ can be understood from the level diagram shown in figure \ref{fig:SepcExampleGG} b). With the molecule initially in the ground state, the energy $\hbar \nu$ extracted from the trap has to match the energy to fully excite the TLS $(- \hbar \Delta)$ and the additional energy to excite the cavity ($\hbar \Delta_{\rm c}$). In analogy to the discussion given above we distinguish two regimes of $\nabla g$-cooling, namely the resolved sideband regime $\kappa \ll \nu$ (SB)  and the Doppler regime $\kappa \gg \nu$ (D) with the corresponding limits for the steady state occupation number,
\begin{equation}
({\rm SB}):\quad  \langle n \rangle_0  = \left({\kappa \over 2 \nu}\right)^2,\qquad ({\rm D}): \quad  \langle n \rangle_0  = {\kappa \over 2 \nu}.
\end{equation}
Therefore, at zero temperature the final occupation number produced by $\nabla g$-cooling may be discussed in analogy to the cavity-assisted cooling,
with the decay rate $\gamma$ replaced by the photon decay rate $2\kappa$. The sideband resolved regime,  which has been discussed in the optical regime e.g. in reference~\cite{VuleticPRA2001}, seems to be accessible in the present setup.

For finite temperature, $\nabla g$-cooling becomes more involved as can be seen from the spectrum $S_g(\omega)$ plotted in figure \ref{fig:SepcExampleGG}. In the resolved sideband regime, $S_g(\omega)$ now consists of four distinct resonances at $\omega=\pm (\Delta_{\rm c}-\Delta)$ and $\omega=\pm \Delta_{\rm c}$. The resonance at $\omega=\Delta_{\rm c}-\Delta$ corresponds to the process discussed above for zero temperature,
and the resonance at $\omega=-(\Delta_{\rm c}-\Delta)$ corresponds to the analogue process with the molecule initially in the excited state. The two resonances at the cavity detuning $\omega=\pm \Delta_{\rm c}$ describe thermally activated processes which do not involve the external driving field $\Omega$. We interpret the corresponding heating and cooling rates as a cavity mediated thermalization process of the molecular motion, which otherwise would be highly decoupled from the electromagnetic environment. In the weak driving limit we find that the resonances at $\omega=\pm \Delta_{\rm c}$ are of equal height and therefore result in a purely diffusive ($W=0$) dynamics. In other words, due to the large frequency mismatch $\omega_{\rm c} \gg \nu$ implicitly assumed in the derivation of our model, any finite thermal occupation of the cavity mode translates into an effective infinite temperature reservoir for the molecular motion.

In the resolved sideband limit $\kappa \ll \nu$ we minimize re-thermalization rates by choosing either $|\Delta_{\rm c}| \gg |\Delta_{\rm c}-\Delta|$ or $|\Delta_{\rm c}|=0$.
In section \ref{sec:Discussion} we show that  $\nabla g$-cooling at finite temperatures results in a steady state occupation number,
\begin{equation}
\langle n \rangle_0  = N + (N+1) \left({\kappa \over 2 \nu}\right)^2 + \langle n \rangle_{\rm th} ,
\label{eq:NF_D_SBth}
\end{equation}
where $\langle n \rangle_{\rm th} \sim 8 (N+1) \rho_{ee}^0 ( \kappa / \Omega)^2 $ for both choices on $|\Delta_{\rm c}|$. Thus in contrast to cavity-assisted sideband cooling, we find that for $\nabla g$-cooling at finite temperature the final occupation number depends crucially on the driving field $\Omega$. A similar result is obtained for the Doppler regime,
\begin{equation}
\langle n \rangle_0  = {\kappa \over 2\nu} \left[ (2N+1) + \beta_{\rm th} \right],
\label{eq:NF_D_Dth}
\end{equation}
with $\beta_{\rm th} \sim 16 (N+1) \rho_{ee}^0 ( \kappa / \Omega)^2 $ for the same choices on $|\Delta_{\rm c}|$. From equations (\ref{eq:NF_D_SBth}) and (\ref{eq:NF_D_Dth}) we conclude that to employ the mechanism of $\nabla g$-cooling efficiently even at finite temperature we must go beyond the weak driving limit $\Omega\rightarrow 0$ and consider either $\Omega\sim \kappa$ or $\Omega \sim |\Delta|$. In the regime $\Omega\sim \kappa\ll |\Delta|,|\Delta_{\rm c}|$, the cavity is far detuned from the rotational transition frequency and cavity mediated thermalization is suppressed, but it requires enough power to scatter photons from the driving field off-resonantly into the cavity. Alternatively we may consider the regime $\Omega \sim |\Delta|$ such that the driven excited state population $\rho^\Omega_{ee}=\Omega^2/[(4\Delta^2+2\Omega^2+\gamma^2)]$ becomes comparable to $\rho_{ee}^0=N/(2N+1)$ and stimulated cooling transitions eventually dominate over thermally activated heating processes. However, in this regime equation (\ref{eq:NF_D_SBth}) and (\ref{eq:NF_D_Dth}) are no longer valid and we postpone the discussion of $\nabla g$ cooling in the strongly driven regime $\Omega \gg |\Delta |$  to section \ref{sec:Discussion} where we find a significant improvement also for a low driving power $\Omega > \gamma$.

\subsubsection{Cooling by quantum interference.}

As a consequence of interference effects between forces exerted on the molecule by the cavity and the classical microwave field, the spectrum $S(\omega)$ plotted in figure \ref{fig:SpecExample} clearly differs from the naively expected sum of $S_\Omega(\omega)$ and $S_g(\omega)$. In equation (\ref{Eq:S_contr}) we have introduced the term $S_{I}(\omega)$ to account for this difference.

\begin{figure}
\begin{centering}
\includegraphics[width=0.85\textwidth]{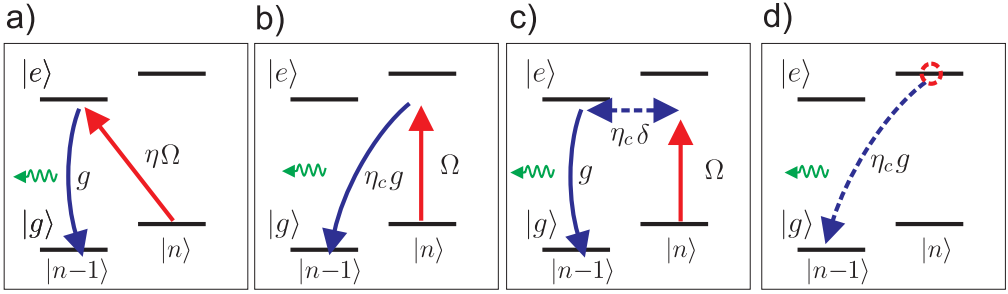}
\caption{a)-c) Different paths which interfere in the cooling process.
To emphasize the underlying coherent interaction we have decomposed the effective decay process in a coherent energy transfer ($g$) and a successive emission of a cavity photon. d) Contribution to the cooling rate which at finite temperature adds incoherently to the other paths. } \label{fig:Ipaths}
\end{centering}
\end{figure}

In figure \ref{fig:Ipaths} we have sketched the three different paths which contribute to the interference effects. In a) the molecule is excited on the red sideband followed by an effective decay via the cavity on the carrier transition. This process is described by $S_\Omega (\omega)$. In b) the molecule is excited on the carrier transition and changes its motional state in the decay process, described by $S_g (\omega)$. The third path sketched in figure \ref{fig:Ipaths} c) might not be obvious in the first place and has the following interpretation. The position dependent coupling  of the TLS to the cavity mode, $g(\hat x)$, in general leads to a position dependent decay constant $\gamma(\hat x)$ as well as a position dependent Stark shift $\delta(\hat x)$. In a semiclassical picture, the oscillating motion of the molecule in the trap therefore translates into modulations of $\gamma$ and $\delta$ at the frequency $\nu$ and results in additional sidebands in the excitation process.  Note that interference between path b) and path c) also occurs for $\eta\rightarrow 0 $ and in general we cannot write $S(\omega)\equiv S_g(\omega)$ even in this limit. However, in section \ref{sec:Discussion} we will argue that in the parameter regimes relevant for $\nabla g$ cooling those interference effects only slightly modify the cooling process. For finite temperature a fourth process, the decay of a thermally excited molecule as sketched in figure \ref{fig:Ipaths} d), adds incoherently to the processes a)-c). Again, this process is proportional to the thermal equilibrium excited state population $\rho_{ee}^0$.

Cooling by quantum interference was discussed and demonstrated in the context of electromagnetically induced transparency~\cite{MorigiPRL2000,RoosPRL2000} and has been proposed for atoms in optical cavities~\cite{CiracPRA1995,ZippilliPRL2005}. In all those schemes cooling relies on a reduction or ideally a cancelation of blue sideband transitions ($S(-\nu)\rightarrow 0$) by interference, while $S(\nu)$ is still finite, which in the ideal case leads to ground state cooling, $\langle n\rangle_0=0$. From figure \ref{fig:SpecExample} we see that in the present system such an exact cancelation of the heating rate appears in the weak driving regime, but is no longer present in the strongly driven system. At finite temperature thermally activated processes reduce the contrast of the interference features, and from the discussion of the individual paths given above we conclude that cooling by quantum interference is only efficient as long as driven cooling  and heating processes $\sim \rho^\Omega_{ee}$ dominate over thermally activated processes $\sim\rho_{ee}^0$. Due to the sensitivity to temperature as well as other system parameters, we do not discuss cooling by quantum interference in much detail in the present work, but in section
 \ref{sec:Discussion} we point out several consequences of interference effects for the combination of CASC and $\nabla g$-cooling. In references \cite{CiracPRA1995,ZippilliPRL2005,ZippilliPRA2005} more details on interference effects in cavity-assisted cooling of atoms can be found for the weak excitation regime ($N=0$, $\Omega\rightarrow 0$).




\section{Theory of Cavity Cooling in the bad cavity limit}\label{sec:Theory}

In the previous overview section we have presented for the case of a weakly driven molecule a physical picture of the individual cooling mechanisms which contribute to the overall cavity-assisted cooling process. For the weak driving regime and for zero temperature, $(\Omega\rightarrow 0, N=0)$ analytic expressions for $S(\omega)$ and the resulting cooling rates have been derived in previous work~\cite{CiracPRA1995,ZippilliPRA2005} on a related system of trapped atoms inside optical cavities. However, for non-zero temperature we have identified thermal cavity-mediated heating processes which degrade the cooling efficiency, and thus in the present setup there is a clear necessity to extend the discussion of cavity cooling schemes to the regime of a strongly driven molecule. In this section we present a derivation of heating and cooling rates which is valid in the bad cavity limit
$g  \ll \kappa$ and allows us to study cavity cooling for a wide range of system parameter.

The starting point for the derivation is the combined cavity-molecule system discussed in section \ref{sec:Overview}, for which the
evolution of the total density operator $\rho(t)$ is determined by the following master equation,
\begin{equation}
\dot \rho \ = \ \mathcal{L}(\rho) \ = \ -i[H_{\rm S} , \rho]+ \mathcal{L}_\kappa(\rho).
\label{Eq:dotrho}
\end{equation}
The system Hamiltonian $H_{\rm S}$ is defined in equations (\ref{eq:HsGen}) and (\ref{Eq:Hs}), and the coupling of the cavity field to a thermal reservoir is taken into account by the term
\begin{equation}
\mathcal{L}_\kappa(\rho) = \kappa (N+1) \ \mathcal{D}[c](\rho) + \kappa N \ \mathcal{D}[c^\dagger](\rho)\,,
\end{equation}
with $\mathcal{D}[c](\rho) = 2 c \rho c^\dagger - \rho c^\dagger c - c^\dagger c \rho $. Starting from the full master equation (\ref{Eq:dotrho}) we proceed in two steps.
First, we make use of the small parameter $g/\kappa$ to eliminate the dynamics of the cavity mode and derive an effective master equation for the internal (TLS) and external degrees of freedom of the molecule. In a second step, the weak coupling between the TLS and the external degrees of freedom, represented by the small parameters $\eta,\eta_{\rm c}$, also allow us to eliminate the TLS and derive an effective master equation describing only the motion of the molecule in the trap.

\subsection{Adiabatic elimination of the cavity mode}
For the elimination of the cavity degrees of freedom we decompose the total Liouville operator $\mathcal{L}$ defined in equation (\ref{Eq:dotrho}) into  three contributions
\begin{equation}
\mathcal{L} = \mathcal{L}_{\rm c} + \mathcal{L}_{\rm m} +  \mathcal{L}_g.
\label{Eq:Liouvillian}
\end{equation}
The first two terms describe the uncoupled dynamics of the cavity mode,
\begin{equation}
\mathcal{L}_{\rm c} (\rho) = -i [(\Delta_{\rm c} - \Delta)c^\dag c , \rho] + \mathcal{L}_\kappa (\rho),
\label{Eq:LcLm}
\end{equation}
and the molecular degrees of freedom,
\begin{equation}
\mathcal{L}_{\rm m} (\rho)= - i [H_{\rm m}, \rho],
\label{eq:HeHi}
\end{equation}
whereas the interaction between the cavity mode and the molecule is described by the last term in equation (\ref{Eq:Liouvillian}),
\begin{equation}
\mathcal{L}_g (\rho) = - i  [g(\hat{x})(\sigma_+ c + \sigma_- c^\dagger), \rho].
\label{Eq:Lg}
\end{equation}
In the decoupled  limit $g\rightarrow 0$ the cavity field evolves independently of the molecule. On a timescale $\kappa^{-1}$ the system relaxes into the state $\rho(t)\approx \rho^0_{\rm c} \otimes \rho_{\rm m}(t)$, with $\rho_{\rm c}^0$ the cavity equilibrium density operator defined by $\mathcal{L}_{\rm c}(\rho_{\rm c}^0)=0$, and the molecular operator $\rho_{\rm m}(t)$ evolving under the action of $\mathcal{L}_{\rm m}$. For a finite coupling $g\ll\kappa$ deviations from the factorized form of $\rho(t)$  are small, but the coupling term $\mathcal{L}_g$ modifies the dynamics of $\rho_{\rm m}(t)$. To proceed we adopt a projection operator technique along the lines of reference \cite{CiracPRA1992} and define the projector
 $ \mathcal{P}\rho = \rho_{\rm c}^0 \otimes {\rm Tr}_{\rm c}\{ \rho\}$ and its orthogonal complement $\mathcal{Q}=(1-\mathcal{P})$. Inserting the decomposition $\rho(t)=\mathcal{P}\rho(t)+\mathcal{Q}\rho(t)$ into equation (\ref{Eq:dotrho}) we obtain the two coupled equations
 \begin{eqnarray}
 \mathcal{P}\dot \rho(t)&=&  \mathcal{P}\mathcal{L}_{\rm m}\mathcal{P}\rho(t) + \mathcal{P}\mathcal{L}_g \mathcal{Q}\rho(t),
 \label{Eq:Pdot}\\
 \mathcal{Q}\dot \rho(t)&=& \mathcal{Q}(\mathcal{L}_{\rm c}+\mathcal{L}_{\rm m}+ \mathcal{L}_g)\mathcal{Q}\rho(t) + \mathcal{Q}\mathcal{L}_g \mathcal{P}\rho(t).
 \label{Eq:Qdot}
 \end{eqnarray}
As the population in the subspace $\mathcal{Q}\rho$ is damped with a rate $\kappa$ which is fast compared to the coupling term $\mathcal{Q}\mathcal{L}_g \mathcal{P} \sim g$, we formally integrate equation (\ref{Eq:Qdot}), insert the result into equation (\ref{Eq:Pdot}) and expand the final expression up to second order in $g$.  For times $t \gg \kappa^{-1}$ we end up with an effective master equation for the molecule density operator  $\rho_{\rm m}(t)={\rm Tr}_{\rm c}\{\mathcal{P} \rho(t)\} $ which is given by
\begin{equation}
\fl
\dot \rho_{\rm m} (t)= -i[H_{\rm m},\rho_{\rm m}(t)] + \int_{0}^\infty d\tau \, {\rm Tr}_{\rm c}\{ \mathcal{L}_{g}\mathcal{Q}e^{(\mathcal{L}_{\rm c}+\mathcal{L}_{\rm m})\tau}\mathcal{Q}   \mathcal{L}_{g} e^{-\mathcal{L}_{\rm m}\tau} (\rho_{\rm m} (t)\otimes \rho_{\rm c}^0 ) \}.
\label{Eq:MEmol}
\end{equation}
Strictly speaking the effective master equation (\ref{Eq:MEmol}) is valid in the limit $g\sqrt{N+1}/\kappa \ll 1$, but we expect it to hold to a good approximation also for a slightly relaxed condition, $g\lesssim \kappa$.

The second term in equation (\ref{Eq:MEmol}) describes the effect of the cavity on the molecular dynamics. We evaluate this term by inserting the definitions of $\mathcal{L}_{\rm c}$, $\mathcal{L}_{\rm m}$ and $\mathcal{L}_g$ given in equations (\ref{Eq:LcLm})-(\ref{Eq:Lg}) and obtain an effective master equation on the form (see also~\cite{CiracPRA1992})
\begin{equation}
\fl
\dot \rho_{\rm m} = -i[H_{\rm m},\rho_{\rm m}]+ \left[(N+1) \left(T \rho_{\rm m} S^\dag - S^\dag T\rho_{\rm m} \right) + N \left( T^\dag \rho_{\rm m} S -S T^\dag \rho_{\rm m} \right) + H.c.\right].
\label{Eq:rho_m}
\end{equation}
Here we introduced the abbreviations $S= g(\hat x)\sigma_- $ and
\begin{equation}
T=\int_0^\infty d\tau\, e^{-i(\Delta_{\rm c}-\Delta)\tau}\, e^{-\kappa \tau} \, S(-\tau)\,,
\qquad S(t)=e^{iH_{\rm m} t}S e^{-iH_{\rm m} t}.
\label{Eq:T}
\end{equation}
In summary we have derived an effective master equation (\ref{Eq:rho_m}) for the molecule, where the first term describes the bare evolution, and the second term describes the influence of the cavity on the dynamics. We want to point out that the latter term in general has a complicated form, since the operator $T$ (\ref{Eq:T}) depends on the bare molecule Hamiltonian $H_{\rm m}$. Only in the infinite bandwidth limit, where $\kappa$ is large compared to $\Delta,\Omega$ and $\nu$, will equation (\ref{Eq:T}) reduce to the simple form $T=S/(\kappa-i\Delta_{\rm c})$.

\subsection{Effective molecule dynamics in the Lamb-Dicke regime}
To proceed with the derivation, we now consider the master equation (\ref{Eq:rho_m}) in the Lamb-Dicke limit $\eta,\eta_{\rm c} \ll 1$
where internal and external degrees of freedom interact weakly.  We expand the position dependent Rabi frequencies $\Omega(\hat x)$ and $g(\hat x)$ up to first order in the Lamb-Dicke parameters, i.e.,
\begin{equation}
\Omega(\hat x)\simeq\Omega[1+\eta(a+a^\dag)] \qquad {\rm and} \qquad S\simeq g\sigma_-[1+\eta_{\rm c} (a+a^\dag)]
\label{Eq:S_LD}.
\end{equation}
Under the same approximation the operator $T$ gets the form
\begin{equation}
T\simeq g\left[ \Sigma_-  +  \eta_{\rm c} \Sigma_-(\nu) a  +  \eta_{\rm c} \Sigma_-(-\nu) a^\dag  \right],
\label{Eq:T_LD}
\end{equation}
with $\Sigma_- (\nu), \Sigma_+ (\nu) = \Sigma_-^\dagger (-\nu)$ effective TLS operators defined in (\ref{Eq:SigmaInt}).
Up to second order in the Lamb-Dicke parameters we obtain the master equation
\begin{equation}
\dot\rho_{\rm m}(t) \ \simeq \ (\mathcal{L}_0 +\mathcal{L}_1+\mathcal{L}_2) \ \rho_{\rm m}(t)\,,
\label{Eq:rho_m_exp}
\end{equation}
where the subscripts indicate the respective order in the Lamb-Dicke parameters $\eta,\eta_{\rm c}$. The zeroth order term is given by
\begin{equation}
\mathcal{L}_0=\mathcal{L}_{\rm I}+\mathcal{L}_{\rm E} ,
\label{Eq:L0}
\end{equation}
where $\mathcal{L}_{\rm E}(\rho)=-i\nu[a^\dag a,\rho]$ describes the uncoupled external dynamics of the molecule, and $\mathcal{L}_{\rm I}$ describes
the effective dynamics of the uncoupled TLS,
\begin{eqnarray}
\mathcal{L}_{\rm I}\rho_{\rm m} \ = \ -i[ H_{\rm I} , \rho_{\rm m} ] \ &+& \ g^2(N+1)\Big[\Sigma_-\rho_{\rm m} \sigma_+-\sigma_+ \Sigma_-\rho_{\rm m} + H.c. \Big]\nonumber \\
\fl
 \ &+& \ g^2N \Big[\Sigma_+\rho_{\rm m} \sigma_--\sigma_- \Sigma_+\rho_{\rm m}+ H.c.\Big]\,.
\label{Eq:LI}
\end{eqnarray}
Explicit expressions for $\mathcal{L}_1$ and $\mathcal{L}_2$ are listed in \ref{app:spectrum}.
Let us for the moment consider the effective dynamics of the TLS (\ref{Eq:LI}), with the
corresponding  Bloch equations for the set of operators $\{\sigma_x , \sigma_y  ,  \sigma_z \}$ of the form
\begin{equation}
\fl
\left(\begin{array}{c}
\langle\dot \sigma_x\rangle \\ \langle\dot \sigma_y\rangle \\ \langle\dot \sigma_z\rangle
\end{array}
\right) =
\left(
\begin{array}{ccc}
- (\tilde \gamma_N - \gamma_x)/ 2 & \Delta + \tilde \delta - \delta_x & 0 \\
- (\Delta + \tilde \delta + \delta_y) & - (\tilde \gamma_N + \gamma_y)/ 2 & - \Omega \\
\Omega_x  & \Omega + \Omega_y & -\tilde \gamma_N
\end{array}
\right)
\left(\begin{array}{c}
\langle\sigma_x\rangle \\ \langle\sigma_y\rangle \\ \langle\sigma_z\rangle
\end{array}
\right) -\left(
\begin{array}{c}
\Gamma_x \\
\Gamma_y \\
\tilde \gamma
\end{array}
\right).
\label{bloch}
\end{equation}
Here $\tilde \gamma_N=(2N+1)\tilde \gamma$ and the remaining parameters are defined in \ref{app:bloch}. In the discussion in section \ref{sec:Discussion} we will mainly focus on the limit $\Omega\ll \kappa$ where the parameters $\delta_{x,y},\gamma_{x,y},\Omega_{x,y}$ and $\Gamma_{x,y}$ are small and equation (\ref{bloch}) reduces to the normal Bloch equations for a two level system coupled to a thermal bath. Further the stark shift $\tilde \delta$ and the effective decay rate $\tilde \gamma$ reduce to the familiar expressions
\begin{equation}
\tilde \delta \rightarrow \delta = (2N+1) \frac{g^2 \Delta_{\rm c}}{\kappa^2+\Delta_{\rm c}^2},
\qquad
\tilde \gamma \rightarrow \gamma = \frac{2g^2 \kappa}{\kappa^2+\Delta_{\rm c}^2}\,.
\end{equation}
For the purpose of readability, we will use the simplification $\Delta + \tilde \delta \rightarrow \Delta$ in the following.
Finally note that when the driving strength $\Omega$ is comparable to the cavity linewidth $\kappa$, unless $|\Delta|, |\Delta_{\rm c}| \gg \kappa$ there will be modifications of $\delta$ and $\gamma$.
In addition, the non-standard terms in the Bloch equations (\ref{bloch}) introduce new features in the TLS dynamics which modify the force spectrum $S(\omega)$ and in the end the cooling dynamics. A detailed discussion of the effective dynamics of a strongly driven TLS inside a cavity is given in \cite{CiracPRA1991}.

\subsection{Adiabatic elimination of the internal degrees of freedom}

Starting from equation (\ref{Eq:rho_m_exp}) we further eliminate the internal degrees of freedom in a similar fashion as the above elimination of the cavity mode, following the procedure in \cite{CiracPRA1992}. First we notice that the zeroth order term ${\mathcal L}_0$ in the expansion (\ref{Eq:rho_m_exp}) describes the uncoupled dynamics of the internal and external degrees of freedom, whereas the higher order terms ${\mathcal L}_{1,2}$ couple the external and internal dynamics and therefore introduce forces acting on the molecule motion. Provided the coupling is weak compared to the energy scales defining the TLS dynamics,
$\eta \Omega , \eta_{\rm c} g^2/\kappa \ll \gamma_N$, the state of the TLS will deviate only slightly from the equilibrium state due to the interaction, and
 the dynamics of the TLS can be eliminated. We define a projection operator $\mathcal{P}_\mu$,
\begin{equation}
\mathcal{P}_\mu(\rho_{\rm m})= \rho_{\rm I}^0 \otimes \sum_n |n\rangle \langle n|\,{\rm Tr}_{\rm I}\{\langle n|\rho_{\rm m}|n\rangle\}\,,
\end{equation}
with $\rho_{\rm I}^0$ the internal equilibrium state defined by $\mathcal{L}_0(\rho_{\rm I}^0)=0$. Second order perturbation theory in the Lamb-Dicke parameters
$\eta, \eta_{\rm c} \ll 1$ results in a master equation for the external operator $\mu(t)={\rm Tr}_{\rm I}\{\mathcal{P}_\mu \rho_{\rm m}(t)\}$, which describes the evolution of the population of the individual trap levels. We obtain
\begin{equation}
\fl
\dot \mu(t) \ = \ \int_0^\infty d\tau\, {\rm Tr}_{\rm I}\{\mathcal{P}\mathcal{L}_1e^{\mathcal{L}_0\tau}  \mathcal{Q}\mathcal{L}_1(\rho_{\rm I}^0\otimes \mu(t))\}
\ + \ {\rm Tr}_{\rm I}\{\mathcal{P}_\mu \mathcal{L}_2(\rho_{\rm I}^0\otimes \mu(t))\}.
\label{eq:dotmu}
\end{equation}
Inserting the definitions of $\mathcal{L}_{1,2}$ given in \ref{app:spectrum} we end up with a master equation on the form
\begin{equation}
\dot \mu(t) = A_-\mathcal{D}[a](\mu(t)) +  A_+\mathcal{D}[a^\dag](\mu(t))\,,
\end{equation}
and recover the cooling equation (\ref{Eq:dotn}) given in section \ref{sec:Overview} with $W=(A_--A_+)$ and $\langle n\rangle_0=A_+/W$.
The heating and cooling coefficients $A_\pm$ which follow from an evaluation of equation (\ref{eq:dotmu})  can be divided into three contributions,
\begin{equation}
A_\pm= S_{\Omega}(\mp\nu) + S_g(\mp \nu) + S_{I}(\mp \nu)\,,
\label{Eq:Apm}
\end{equation}
where the physical motivation for the decomposition (\ref{Eq:Apm}) has been presented in section \ref{sec:Overview}.
The first contribution to the force spectrum, $S_\Omega(\omega)$, follows from the first term of equation (\ref{eq:dotmu}) in the limit $\eta_{\rm c}\rightarrow 0$ and is defined as
\begin{equation}
S_\Omega (\omega) = \frac{\eta^2\Omega^2}{2} {\rm Re}  \int_0^\infty d\tau \, {\rm Tr }_{\rm I} \{\sigma_x e^{\mathcal{L}_0\tau} (\sigma_x\rho_{\rm I}^0)  \} e^{i\omega \tau}\,.
\label{Eq:Somega}
\end{equation}
The second term, $S_g(\omega)$, follows from the second term in equation (\ref{eq:dotmu}) and is given  by
\begin{equation}
 S_g (\omega) = 2 \eta_{\rm c}^2 g^2 \ {\rm Re} \big\{(N+1) \langle \sigma_+ \Sigma_-(\omega) \rangle_{\rm I} + N \langle \sigma_-\Sigma_+(\omega)\rangle_{\rm I} \big\},
 \label{Eq:Sg}
\end{equation}
with effective TLS operators $\Sigma_\pm (\nu)$ given by (\ref{Eq:SigmaMinus}).
Finally, remaining contributions from the first term in equation (\ref{eq:dotmu}) are found by making the Ansatz,
\begin{equation}
\fl
\mathcal{L}_1(\rho_{\rm I}^0 \otimes \mu )= \mathcal{K}_+(\rho_{\rm I}^0) \otimes a\mu +   \mathcal{K}_-(\rho_{\rm I}^0) \otimes a^\dag \mu +  \mathcal{K}^\dag_-(\rho_{\rm I}^0) \otimes \mu a  +  \mathcal{K}^\dag_+(\rho_{\rm I}^0)\otimes \mu a^\dag  \,.
\label{Eq:L1K}
\end{equation}
The result is
summarized by $S_{I}(\omega)$ defined as
\begin{equation}
S_{I} (\pm \nu) = 2 {\rm Re} \int_0^\infty d\tau \, {\rm Tr }_{\rm I} \{ \mathcal{K}^\dag_\pm ( e^{\mathcal{L}_0\tau} \mathcal{K}_\pm (\rho_{\rm I}^0) ) \} e^{\pm i\nu \tau}\, -S_\Omega (\pm \nu) ,
\label{Eq:SI}
\end{equation}
with the action of the superoperators $\mathcal{K}_\pm$ given by
\begin{eqnarray}
\fl
\mathcal{K}_\pm(\rho) = - i  \frac{\eta\Omega}{2}\sigma_x \rho
&+ & \eta_{\rm c} g^2 (N+1)[ \sigma_- \rho \Sigma_+  + \Sigma_-(\pm\nu)\rho \sigma_+ -\sigma_+\Sigma_-(\pm\nu)\rho -\sigma_+ \Sigma_- \rho]  \nonumber \\
 &+& \eta_{\rm c} g^2 N [ \sigma_+ \rho \Sigma_-  + \Sigma_+(\pm\nu)\rho \sigma_- -\sigma_-\Sigma_+(\pm\nu)\rho -\sigma_- \Sigma_+ \rho]\,.
 \label{Eq:Koperator}
\end{eqnarray}

In conclusion, we have expressed heating and cooling rates $A_\pm$ in terms of two-point correlation functions and steady state expectation values of a TLS whose evolution is described in terms of the effective Bloch equations (\ref{bloch}). The results are valid for $g\sqrt{N+1}\ll\kappa$ and $\eta \Omega \ll \gamma_N $, $\eta_{\rm c} \ll 1$ but at this stage no further assumptions have been made. However, in the following we mainly concentrate on the case $\Omega \ll \kappa$ or on the regime $\Omega\sim \kappa$, $|\Delta_{\rm c}|,|\Delta|\gg \Omega$ where the Bloch equations reduce to the standard form (see \ref{app:bloch}). Using the quantum regression theorem and the steady state solution of equation (\ref{bloch}) we outline the derivation of analytic expressions for $S_\Omega(\omega)$ and $S_g(\omega)$ in \ref{app:S} and \ref{app:g}. Analytic expressions for $S_{I}(\omega)$ could in principle be derived along the same lines but in general do not have simple enough form to provide further insight. Therefore we will rather use numerical calculations in combination with previously derived results for the weak excitation limit~\cite{ZippilliPRA2005} as a basis for the discussion of the interference terms.

%
%

\section{Discussion}
\label{sec:Discussion}

In this section we discuss cavity-assisted cooling of polar molecules both in the weak and strong driving regime, and base the discussion on the result for $S(\omega)$ derived in section \ref{sec:Theory}. For a physical understanding we again find it useful to discuss cavity assisted sideband cooling and $\nabla g$ cooling independently in a first step. In a second step we will then study the general case where both mechanisms as well as interference effects are taken into account. Analytical results for cooling rates and final occupation numbers for the different parameter regimes are summarized at the end of this paper in table~\ref{tab:summary}.

\subsection{Cavity assisted sideband cooling}
\label{ssec:CASC}
We start the discussion with the mechanism of cavity assisted sideband cooling (CASC) where the corresponding cooling rate $W_\Omega=S_\Omega(\nu)-S_\Omega(-\nu )$ and  steady state occupation $\langle n \rangle_{0,\Omega} = S_\Omega (-\nu)/W_\Omega$ depend on the spectrum $S_\Omega(\omega)$ defined in equation (\ref{Eq:Somega}). Following the calculations outlined in \ref{app:S} we obtain for the spectrum in the weak driving limit, $\Omega \ll \gamma_N$,
\begin{equation}
S_\Omega (\omega) = {\eta^2 \Omega^2 \over 4} \left[ {\rho_{gg}^0 \gamma_N \over (\omega + \Delta)^2 + \gamma_N^2/4}
+ {\rho_{ee}^0 \gamma_N \over (\omega - \Delta)^2 + \gamma_N^2/4} \right].
\label{eq:SomegaWeak}
\end{equation}
For a fixed ratio $\nu/\gamma_N$ we optimize $\langle n\rangle_{0,\Omega}$ with respect to the detuning $\Delta$, where obviously $\Delta$ must be negative to promote cooling. For the two limits of interest, the resolved sideband limit (SB) $\nu \gg \gamma_N$ and the Doppler limit (D) $\nu \ll \gamma_N$, we find $\Delta = - \nu$ and $\Delta = -\gamma_N/2$ respectively. The final occupation number is given by the respective expressions (\ref{eq:NfSB}) and (\ref{eq:NfD}) in section \ref{sec:Overview}, and the corresponding cooling rates are given by,
\begin{equation}
({\rm SB}): \quad W_\Omega = {\eta^2 \Omega^2 \over (2N+1)\gamma_N}, \qquad
({\rm D}): \quad W_\Omega = {\eta^2 (2 \nu / \gamma_N) \Omega^2  \over (2N+1)\gamma_N} .
\label{eq:Wweak}
\end{equation}
Although in the weak driving limit the final occupation number is independent of $\Omega$, the cooling rate scales as $W_\Omega\sim \Omega^2$ and under the validity of equation (\ref{eq:Wweak}) it is bounded be $W_\Omega < \eta^2 \gamma$.  In competition with the cavity-mediated thermalization process of the molecule, discussed in section \ref{sec:Overview}, it is therefore necessary to study sideband cooling beyond the weak driving regime.

\begin{figure}
\begin{centering}
\includegraphics[width=0.5\textwidth]{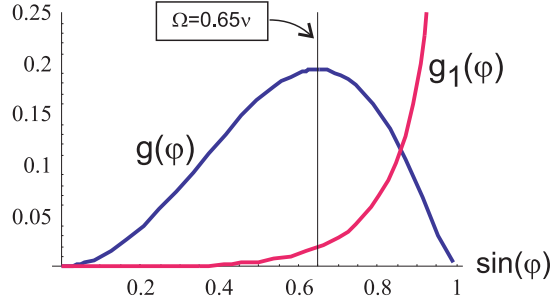}
\caption{Dimensionless cooling rate ${\rm g}(\varphi)$ and final occupation number ${\rm g}_1 (\varphi)$ for CASC in the resolved sideband limit,
as functions of $\sin\varphi = \Omega/\nu$. The vertical line marks the
optimal drive strength $\Omega = 0.65 \nu$ which maximizes the cooling rate.\label{fig:g-functions}}
\par\end{centering}
\end{figure}

Let us now consider an arbitrary $\Omega$, but focus on CASC in the sideband resolved limit
$\gamma_N/\nu \ll 1$. As shown in \ref{app:S}, the spectrum now exhibits a three-peak structure and reads,
\begin{equation}\label{Eq:OmegaSpectrumStrong}
S_\Omega (\omega)\simeq  {\eta^2 \Omega^2 \over 4} \left[ \frac{\alpha_+\bar\gamma }{(\omega-\bar \Delta)^2+\bar\gamma^2/4} +
\frac{\alpha_0 \bar\gamma_0 }{\omega^2+\bar\gamma_0^2/4} + \frac{\alpha_- \bar\gamma}{(\omega+\bar \Delta)^2+ \bar\gamma^2/4} \right]\,,
\end{equation}
with $\bar \Delta=\sqrt{\Omega^2+\Delta^2}$. For sideband cooling we are only interested in the peaks at nonzero frequency. The relative heights of the resonances, $\alpha_\pm$, and the widths $\bar\gamma , \bar\gamma_0$ which are proportional to $\gamma_N$ are defined in \ref{app:S} and are functions of $\Omega$, $\Delta$ and $N$. For $W_\Omega$ to be positive we require a red detuned microwave field $\Delta <0 $ and cooling is optimized for $\nu=\bar \Delta$.
Under those conditions we obtain
\begin{equation}
W_\Omega \ = \ {\eta^2 \Omega^2 \over 4} \left( {\alpha_+ - \alpha_- \over
\bar\gamma } \right)
\ = \ \gamma \left({ \eta \nu \over\gamma_N}\right)^2 {\rm g} (\varphi),
\label{W_S}
\end{equation}
with ${\rm g}(\varphi) = 4 \sin^2 \varphi |\cos^3 \varphi|/(4 - \sin^4 \varphi)$, and
\begin{equation}
\langle n \rangle_{0,\Omega} \ = \ {\alpha_- \over \alpha_+ - \alpha_- } + \mathcal{O}(\gamma_N^2/\nu^2)
 \ \simeq \ N  + (2N+1) {\rm g}_1 (\varphi) ,
\end{equation}
with ${\rm g}_1 (\varphi) = (1 - |\cos\varphi|)^2/(4|\cos\varphi|)$. The functions ${\rm g}(\varphi)$ and ${\rm g}_1(\varphi)$ are plotted in figure \ref{fig:g-functions} and depend on the angle $\varphi$ defined as $\sin(\varphi)=\Omega/\bar\Delta$. We find that the cooling rate increases with $\Omega$ up to $\Omega \sim |\Delta|$ but then goes to zero for $\Omega\gg |\Delta|$. It has a maximum value of ${\rm g}(\varphi_0)\simeq 0.2$ at $\sin\varphi_0 = \Omega_0/\bar\Delta \simeq 0.65$, and we want to point out that this holds independently of the temperature $N$. The final steady state occupation number increases monotonically with increasing $\Omega$; ${\rm g}_1(\varphi_0) \simeq 0.02$ and ${\rm g}_1 \rightarrow \infty$ for $\Omega \gg |\Delta|$.

We conclude that driving moderately, $\Omega \sim |\Delta| \sim \nu/2$, increases the cooling rate tremendously while only marginally increasing the final occupation number. We point out that the validity of the adiabatic elimination constraints the parameters, $\eta \nu \sim \eta \Omega < \gamma_N$,  and thus the expected upper limit $W \leq \gamma$ still holds. Nevertheless, this bound is by a factor $1/\eta^2$ higher than in the weak driving case.
\subsection{$\nabla g$ cooling}
\label{subsec:nablaGdisc}
Let us now move on to discuss the mechanism of $\nabla g$ cooling, where the cooling rate $W_g=S_g(\nu) - S_g (-\nu)$  and the final occupation number $\langle n \rangle_{0,g} = S_g (-\nu)/W_g$ are given in terms of the spectrum $S_g (\omega)$ defined in equation (\ref{Eq:Sg}).  In \ref{app:g} we show that for finite temperature $N\neq 0$ and arbitrary $\Omega \ll \kappa$ or $\Omega \sim \kappa \ll |\Delta|,|\Delta_{\rm c}|$ the structure of $S_g(\omega)$ (\ref{Spectrum_D}) is quite involved and generally exhibits 6 resonances centered at frequencies
$\omega = \pm (\Delta_{\rm c}-\Delta)$ and
$\omega = \pm (\Delta_{\rm c}-\Delta \pm \bar \Delta)$. Note that we assume $\gamma_N \ll \bar\Delta  $, which holds naturally when we later combine $\nabla g$ cooling with CASC. As the height of two of the resonances, at $\omega = \pm (\Delta_{\rm c}-\Delta - s\bar \Delta)$ where $s = {\rm sgn}(\Delta)$, scales as $ \sim
(\Omega/\Delta)^4$ we can for weak driving $\Omega \ll |\Delta|$ neglect those contributions and obtain the spectrum
\begin{equation}
\fl
S_g (\omega)  = {\eta_{\rm c}^2 g^2 \over 2}  \Big[ \
 {\kappa (N+1)(\Omega / \Delta)^2  \over \kappa^2 + (\omega - \Delta_g )^2 }
\ + \ { \kappa N  (\Omega / \Delta)^2 \over \kappa^2 + (\omega + \Delta_g )^2 }
\ + \ \sum_{s=\pm} { 4 \kappa (N +1) \rho_{ee}^0  \over \kappa^2 + (\omega + s\Delta_{\rm c} )^2 } \ \Big].
\label{eq:SgWeak}
\end{equation}
We note that only the driven terms in equation (\ref{eq:SgWeak}) contribute to the cooling rate $W_g$, which for optimized detuning $\Delta_g = \Delta_{\rm c} - \Delta = \kappa$ in the Doppler limit and $\Delta_{\rm c} - \Delta = \nu$ in the sideband limit, respectively, are given by
\begin{equation}
({\rm SB}): \quad W =  {\eta_{\rm c}^2 g^2 \over 2 \kappa } \left( {\Omega \over \Delta} \right)^2 \qquad
({\rm D}): \quad W = {\eta_{\rm c}^2 g^2  \over 2 \kappa }\frac{\nu}{\kappa} \left({ \Omega \over \Delta } \right)^2.
\label{Eq:Wgg}
\end{equation}
The last term in (\ref{eq:SgWeak}) is symmetric in frequency and thus corresponds to a purely diffusive process which is present only when the thermal excited state population $\rho_{ee}^0$ is nonzero. For finite temperature there is thus a competition between the driven cooling processes proportional to $\bar\rho_{ee} \sim \Omega^2/\Delta^2$ and thermal diffusive processes. In the resulting expressions for the final occupation number $\langle n \rangle_{0,g}$, which are given in equations (\ref{eq:NF_D_SBth}) and (\ref{eq:NF_D_Dth}), corrections due to the additional diffusion process appear in the quantities $\beta_{\rm th}$ and $\langle n\rangle_{\rm th}$ which in general are given by
\begin{equation}
\fl
\langle n \rangle_{\rm th} = 4 \kappa^2 (N+1 ) \rho_{ee}^0
\left( {(\Delta/\Omega)^2 \over \kappa^2 + (\Delta_{\rm c} + \Delta)^2} + {(\Delta/\Omega)^2 \over \kappa^2 + \Delta^2}\right) ,\qquad
\beta_{\rm th} = {16  \kappa^2 (N+1) \rho_{ee}^0 (\Delta / \Omega)^2 \over \kappa^2 + \Delta_{\rm c}^2} .
\end{equation}
As already mentioned in section \ref{sec:Overview}, to reduce $\beta_{\rm th}$  and $\langle n\rangle_{\rm th}$ there are two different strategies. We can either stay in the far detuned regime $|\Delta_{\rm c} | \sim |\Delta | \gg \kappa, \Omega$ and increase the driving power $\Omega \sim \kappa$, or alternatively, use a resonant driving field $|\Delta |\rightarrow  0$. In the first case we are still in the far detuned regime, $\Omega \ll |\Delta|$, and equation (\ref{eq:SgWeak}) is still valid, even for $\kappa \sim \Omega$. For a resonantly driven molecule we however need to reevaluate $S_g(\omega)$, as we will do in the following.

We now consider the strong driving regime $\Omega \gg |\Delta |,\gamma_N$ where the rotational transition of the molecule is saturated; $\rho_{ee} \rightarrow 1/2$. In this limit the general spectrum (\ref{Spectrum_D}) is of the form,
\begin{eqnarray}
S_g (\omega) &=&
{\eta_{\rm c}^2  g^2\over 4 }
\left[ \frac{2\kappa (N+1) }{\kappa^2+ [\omega -(\Delta_{\rm c}-\Delta)]^2 } +
\sum_{s=\pm 1}\frac{\kappa (N+1)}{\kappa^2+ [\omega - (\Delta_{\rm c} + s \Omega)]^2 }
 \right]
\nonumber \\
&+&  {\eta_{\rm c}^2 g^2 \over 4}   \left[
\frac{2\kappa N }{\kappa^2+ [\omega +(\Delta_{\rm c}-\Delta)]^2 }
+  \sum_{s=\pm 1} \frac{\kappa N}{\kappa^2+ [\omega +(\Delta_{\rm c} + s \Omega )]^2 }
\right].
\label{Spectrum_D_strong}
\end{eqnarray}
We find that for a saturated molecule at moderate driving strength $|\Delta |,\gamma_N < \Omega < \kappa$, where the second condition is already implicitly assumed in the derivation of equation (\ref{Spectrum_D_strong}), the three peaks overlap almost completely and thus to a very good approximation we can consider all resonances
to be centered around $ \Delta_{\rm c}$. We then obtain the two peak structure
\begin{eqnarray}
S_g (\omega) &\simeq&
{\eta_{\rm c}^2  g^2}
\left[ \frac{\kappa (N+1) }{\kappa^2+ (\omega -\Delta_{\rm c})^2 }  +\frac{\kappa N }{\kappa^2+ (\omega +\Delta_{\rm c})^2 }
 \right]\,,
\label{Spectrum_D_strong_simple}
\end{eqnarray}
in analogy to the spectrum $S_\Omega(\omega)$ (\ref{eq:SomegaWeak}) in the weakly driven regime. In the fully saturated regime the final temperatures achievable by $\nabla g$ cooling are given by
\begin{equation}
\fl \qquad
({\rm SB}): \quad \langle n \rangle_{0,g} = N+ (2N+1) {\kappa^2 \over 4 \nu^2},
\qquad
({\rm D}): \quad \langle n \rangle_{0,g} = (2N+1) {\kappa \over 2 \nu},
\end{equation}
and exhibit a similar dependence as in the case of CASC but with $\gamma$  replaced by $2\kappa$. However, for a finite environment temperature $\langle n\rangle_{0,g}$ grows in both regimes only linearly with $N$. The cooling rates are independent of $N$ and approach the optimal values of
\begin{equation}
({\rm SB}): \quad W_g = {\eta_{\rm c}^2 g^2  \over \kappa},
\qquad
({\rm D}): \quad W_g = { \eta_{\rm c}^2 g^2 ( \nu / \kappa ) \over \kappa} .
\label{Eq:WggStrong}
\end{equation}

In conclusion we find that the apparent inefficiency of $\nabla g$-cooling at finite temperatures can be overcome either in the regime where the cavity is far detuned from the rotational transition and the driving strength $\Omega$ is comparable to the cavity linewidth $\kappa$, or in the fully saturated regime $\Delta=0$, $\Omega\gg\gamma$. In the later case the cooling rates are optimized and the resulting final temperatures grow only linearly with $N$.

\subsection{Combining cavity-assisted microwave cooling and $\nabla g $ cooling.}

Until now we have only considered the idealized limits $\eta\rightarrow 0$ and $\eta_{\rm c}\rightarrow 0$ where either the mechanism of cavity-assisted sideband/Doppler cooling or $\nabla g$-cooling dominate the cooling process. In this section we focus on the more realistic case where both Lamb-Dicke parameters are of the same order and discuss the combination of the cooling mechanisms, being aware that in general $W\neq W_\Omega+W_g$ due to interference effects. In view of the tight confinement required to trap the molecule close to the surface of the chip, the following discussion is restricted to the resolved sideband regime $\gamma_N < \nu$.

From our analysis so far, we conclude that under ideal conditions cavity cooling of polar molecules is dominated by the mechanism of CASC, which provides the highest cooling rates and results in the lowest final temperature both in the weak and strong driving regime.
However, although CASC is the more effective of the two mechanisms, we expect it also to be more sensitive to imperfections in the system, which in a chip based setup might be harder to avoid than in e.g. a large ion trap. Therefore, in the following we are particularly interested in the robustness of cavity cooling under non-ideal trap conditions. In order to discuss different types of imperfections like anharmonicity or a state dependent trapping potential within a general and simple model, we proceed as follows. We assume that the system parameters are optimized with respect to the expected trapping frequency $\nu$ which differs from the real trapping frequency $\nu_r=\nu+\delta\nu$ by a shift $\delta \nu$, and then we study the dependence of the cooling parameters $W(\delta \nu)$ and $\langle n \rangle_0 (\delta\nu)$.
For an anharmonic trap we can then interpret $W_n=W(\delta\nu_n)$ as a level dependent cooling rate, assuming a frequency spacing of the trap levels $\nu_n=\nu+\delta\nu_n$ with $\delta \nu_0=0$. Small deviations from a state-independent trapping potential (e.g. assuming $\nu$ is the trap level spacing as seen by state $|g\rangle$ and $\nu + \delta\nu$ as seen by state $|e\rangle$) translate into a level dependent detuning $\Delta_n\rightarrow \Delta -n \delta\nu$. Since for sideband cooling, which is most sensitive to imperfections, a shift of the transition frequency and a shift of the trapping frequency affect cooling the same way
we can interpret $W_n=W( n\delta\nu)$ as a level-dependent cooling rate as well.

\subsubsection{Weak saturation regime.}

\begin{figure}
\begin{centering}
\includegraphics[width=0.9\textwidth]{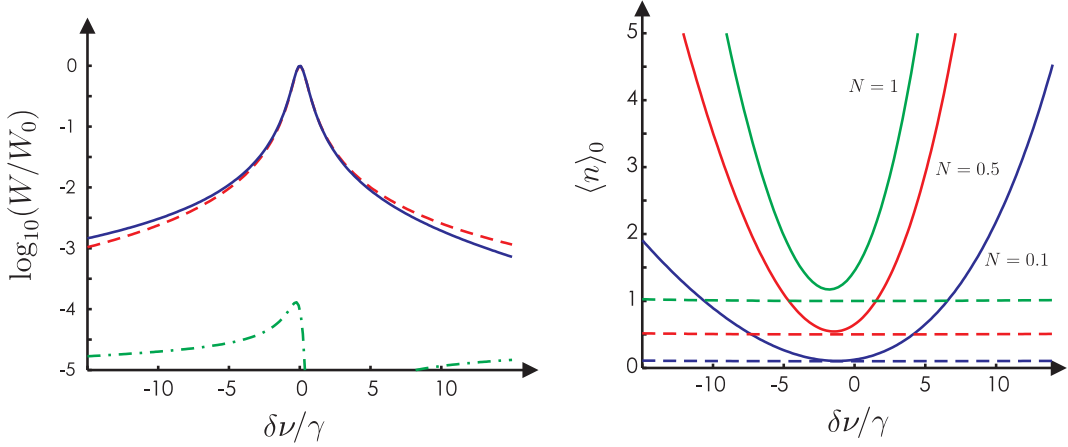}
\caption{Cavity cooling in the weak saturation regime for the parameters (all in units of $\nu$) $\Omega=0.1$, $\Delta=-1$, $\kappa=4$, $\Delta_{\rm c}=3$, $g=0.2$ and $\eta=\eta_{\rm c}$. a) Cooling rate $W(\delta \nu)$ in units of $W_0=\eta^2\Omega^2/\gamma$. The dashed (dashed-dotted) line indicates the corresponding result for $\eta_{\rm c}=0$ ($\eta=0$).  b) Steady state occupation number $\langle n\rangle_0(\delta \nu)$ for N=0.1 (blue), N=0.5 (red) and N=1 (green). Dashed lines for $\eta_{\rm c}=0$.
\label{Fig:CavityCoolingWeak}}
\par\end{centering}
\end{figure}

We first consider the weak saturation regime $\Omega \ll |\Delta|$ and assume the relation $\gamma < \nu  < \kappa $ for the damping parameters. We choose $\Delta=-\nu$ to maximize excitations of the molecule on the red sideband and choose $\Delta_{\rm c}=\kappa-\nu$ with the intention to optimize $\nabla g$-cooling in the Doppler limit. In that case the spectrum $S(\omega)$ is similar to the one plotted in figure \ref{fig:SpecExample} a) in section \ref{sec:Overview}. As an example we plot in figure \ref{Fig:CavityCoolingWeak} the cooling rate $W(\delta \nu)$ and the final occupation number $\langle n\rangle_0 (\delta \nu )$ for a specific set of parameters. We find that in the weak saturation regime  the cooling rate is dominated by the mechanism of CASC and under non-ideal conditions $\delta \nu \neq 0$ it scales as
\begin{equation}
W(\delta \nu)\simeq W_\Omega(\delta\nu)\simeq {\eta^2 \Omega^2 \over (2N+1) \gamma_N } \times
{1 \over 1 + \left( 2 \delta\nu / \gamma_N  \right)^2 }.
\end{equation}
The correction due to the gradients of $g(\hat x)$ lead to an increase of $W(\delta \nu)$ for $\delta \nu < 0$, and a decrease for $\delta \nu >0$. The asymmetry shows that  corrections arise from interference effects in the excitation of the molecule. As expected from the small value of $\rho_{ee}^\Omega \sim \Omega^2/\Delta^2 \ll 1$, the pure mechanism of $\nabla g$-cooling plays essentially no role for the cooling rate of a weakly driven molecule.
However, as discussed above, the mechanical coupling to the cavity introduces an additional source of diffusion which becomes important when the thermally activated excited state population $\rho_{ee}^0$ exceeds $\rho_{ee}^\Omega$. In contrast to pure CASC, where $\langle n\rangle_0(\delta \nu)$ is highly insensitive to $\delta \nu$ (since the heating rate scales like the cooling rate, $A_+ (\delta\nu)\sim 1/[1 + (2\delta\nu/\gamma_N)^2]$), the additional diffusion for $\eta_{\rm c} >0$ scales differently with $\delta\nu$  and will affect the final occupation number. In the limit $\kappa \gg \nu$ we obtain
\begin{equation}
\langle n\rangle_0 (\delta \nu) \ \simeq \ \langle n\rangle_{0,\Omega} \ + \ {W_g \over W_\Omega} \langle n \rangle_{0,g}
\left[ 1 + {4 \over (2N+1)^2} \left( {\delta\nu \over \gamma}\right)^2\right],
\end{equation}
with
\begin{equation}
{W_g \over W_\Omega} \langle n \rangle_{0,g} \ \simeq \ \left( {\eta_{\rm c} \over \eta }\right)^2 { \gamma_N^2 \over 4 \nu^2 }
\left[ (2N+1) + { 8 (N+1) \rho_{ee}^0 \nu^2 \over \Omega^2} \right] .
\end{equation}

In summary we find that for a weakly driven molecule the cavity-assisted microwave cooling is dominated by CASC and under non-ideal conditions $|\delta\nu | > \gamma_N$ the total cooling rate decreases as $W\simeq W_\Omega \sim (\gamma_N/\delta \nu)^2$. For finite temperature the additional diffusion caused by thermal $\nabla g$ processes affects the final occupation number and $\langle n\rangle_0\approx N$ is reached only for $\delta \nu \lesssim \gamma$
and $(\gamma_N/\Omega)^2 \rho_{ee}^0 \lesssim 1$.

\subsubsection{Strong driving regime.}

\begin{figure}
\begin{centering}
\includegraphics[width=0.9\textwidth]{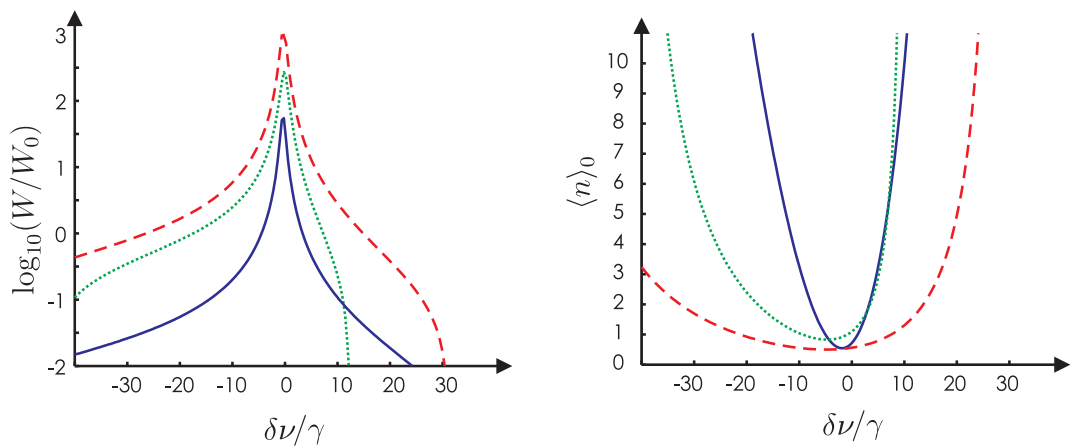}
\caption{Cavity cooling in the strong driving regime for the parameters (all in units of $\nu$) $\bar \Delta=1$ ($\Delta < 0$), $\kappa=4$, $\Delta_{\rm c}=3$, $g=0.2$ and $\eta=\eta_{\rm c}$. a) Cooling rate $W(\delta \nu)$ in units of $W_0=\eta^2\gamma$ for $N=0$ and different values of $\Omega=0.1$ (solid), $0.65$ (dashed) , $0.95$ (dotted).  b) Steady state occupation number $\langle n\rangle_0(\delta \nu)$ for $N=0.5$.
\label{Fig:CavityCoolingGen}}
\par\end{centering}
\end{figure}

We now consider the case of a strongly driven molecule $\Omega\sim|\Delta|$, where according to our previous discussion the cooling rate for CASC, $W_\Omega$, is (close to) optimized.  In figure \ref{Fig:CavityCoolingGen} a) we plot the cooling rate $W(\delta \nu)$ with a fixed $\bar \Delta=\nu$ ($\Delta < 0$) but different values of $\Omega$. Apart from the improvement compared to weaker drive under ideal conditions, we note that $W$ is also less sensitive to deviations $\delta\nu$ for $\Omega\sim|\Delta|$. We attribute this feature partially to power broadening of the resonances, and also to a stronger contribution from interference effects. The asymmetry of the latter improves the robustness of cavity cooling for $\delta \nu <0$ but has a negative effect for $\delta \nu >0$. In figure \ref{Fig:CavityCoolingGen} b) we also plot the final steady state occupation number for $N=0.5$ and different values of $\Omega$.  We see that by increasing $\Omega$ up to $\Omega_0$ the improved robustness of the cooling rate translates into a corresponding robustness of the final temperature with respect to deviations from ideal harmonic trapping conditions. In contrast to $W_\Omega$, the thermal diffusion rates do not increase with $\Omega$, and thus their contribution to the final occupation number is small; the limit $\langle n\rangle_0=N $ can be reached for $(\gamma^2_N/\nu^2)(N+1)\rho_{ee}^0 <1$.
When we increase $\Omega$ beyond the optimal value of $\Omega_0\simeq0.65\nu$ both the efficiency and the robustness of cavity cooling decrease again.

\subsubsection{Fully saturated regime.}

We finally consider the fully saturated regime $\Omega \gg \gamma_N,|\Delta|$ where the effect of CASC vanishes while $\nabla g$ cooling is most effective.
When we assume in addition that $\bar \Delta \sim \Omega \ll \nu$, excitations on the motional sidebands do not play a relevant role any longer and we recover the case of pure
$\nabla g$ cooling as discussed above in section \ref{subsec:nablaGdisc}. According to that discussion, we conclude that cavity cooling in this limit has the same efficiency and robustness as CASC in the weak driving limit, but with the relevant scale $\gamma_N$ replaced by $2\kappa$. As $\gamma_N \ll \kappa$ we trade cooling efficiency for robustness by switching between CASC and $\nabla g$ cooling.
\begin{figure}
\begin{centering}
\includegraphics[width=0.6\textwidth]{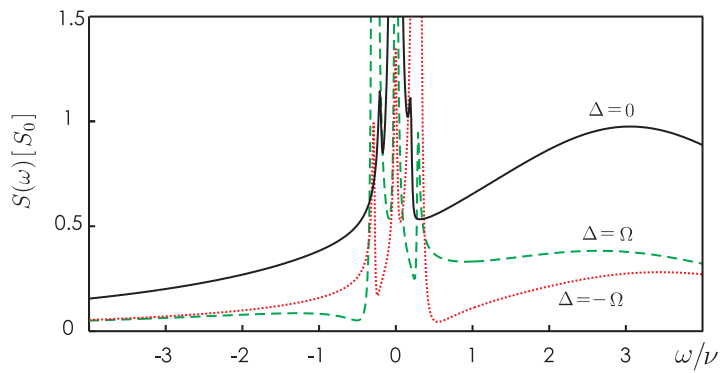}
\caption{Spectrum $S(\omega)$ for $\nabla g$ cooling in the saturated regime. The three different lines show $S(\omega)$ in units of $S_0=\eta_{\rm c}^2 g^2/\kappa$ for $\Delta=0$ (solid), $\Delta=\Omega$ (dashed) and $\Delta=-\Omega$ (dotted). The remaining parameters for this plot have been chosen as (all in units of $\nu$): $g=0.3$, $\kappa=3$, $\Delta_{\rm c}=3$, $\Omega=0.2$ and $\eta=\eta_{\rm c}$. Multiple resonance in the frequency range $|\omega|\leq \sqrt{\Delta^2+\Omega^2}$ are due to excitations of dressed states of the strongly driven TLS.}
\label{Fig:Saturated}
\par\end{centering}
\end{figure}

The arguments presented in the previous paragraph are valid under the assumption $\nu \gg \Omega\gg \gamma_N, |\Delta|$, but these conditions might not always be satisfied in a real experiment. Especially, if the trapping potential is slightly state dependent, the detuning $\Delta$ depends, in a semiclassical picture, on the position of the molecule in the trap. In figure \ref{Fig:Saturated} we plot the total force spectrum $S(\omega)$ for the case  $\Delta=0$ and for $\Delta=\pm \Omega$. Under resonance conditions (note that our definition of $\Delta$ includes the cavity induced Stark shift $\delta$) the spectrum is to a good approximation given by $S_g(\omega)$, assuming $\gamma_N\ll \nu$. For $\Delta$ comparable to $\Omega$ we find that $S(\pm \nu)$ is reduced since the excited state population drops below its maximal value of $1/2$. In addition, for red detuning, $\Delta <0$, destructive interference for $\omega > \bar \Delta$ and constructive interference for $\omega < \bar \Delta$ degrade $\nabla g$ cooling, while a blue detuned driving field, $\Delta > 0$, has just the opposite effect. We conclude that cavity cooling in the saturated regime can be efficiently applied for  a range of parameters fulfilling the condition $\nu > \Omega > \gamma_N, \Delta \geq 0$. Outside this parameter regime, excitations of the molecule on the red and blue motional sidebands as well as destructive interference effects play an important role and $\nabla g$ cooling no longer provides a valid description of the full cooling (or heating) dynamics.

%
%

%
\section{Summary and conclusion}
\label{sec:Summary}

In this paper we have studied  cavity-assisted cooling methods for a single polar molecule, which are based on a strong coupling between two rotational states of the molecule and an on-chip microwave resonator. We have shown that in the Lamb-Dicke limit the cooling dynamics in this system can be understood in terms of two main cooling mechanisms, namely cavity-assisted sideband cooling (CASC) and $\nabla g$ cooling, interference effects between those two mechanisms and a cavity mediated thermalization channel introducing additional heating at finite temperature. Our main conclusions are as follows.

\begin{table}
\caption{\label{tab:summary}
Limits for the final occupation number $\langle n\rangle_0$ and the cooling rate $W$ for the different parameter regimes studied in this paper. For $\nabla g$ cooling  in the weak driving regime, optimized values for $\langle n \rangle_{th},\beta_{th}\sim N(\kappa/\Omega)^2$ are given in section~\ref{sec:Overview}.}
\begin{indented}
\item[]
\begin{tabular}{@{}lll}
\br
   ``CASC''     & \,\,\,\,  sideband limit ($\nu \gg \gamma_N$)   \,\,\,\,   & \,\,\,\, Doppler limit ($\nu \ll \gamma_N$)   \,\,\,\,   \\
\mr
  weak driving  & $\langle n\rangle_0\!=\! N\!+\!(2N\!+\!1)(\gamma_N/4\nu)^2$ & $\langle n\rangle_0\!=\!(2N\!+\!1)\gamma_N/4\nu$ \\
  regime  & $W=\eta^2\Omega^2/((2N\!+\!1) \gamma_N)$  &  $W= 2\eta^2 \Omega^2\nu /((2N\!+\!1) \gamma_N^2)$  \\
\mr
  strong driving & $\langle n\rangle_0\simeq   N+\mathcal{O}(\gamma_N^2/\nu^2)$ &  - \\
  regime  & $W\simeq 0.2  (\eta\nu/\gamma_N)^2 \gamma$  &  - \\
\br
``$\nabla g$ cooling''     &  \, \,\,\, sideband limit ($\nu \gg \kappa$)  \,\,\,  \, &  \,\,\,\,  Doppler limit ($\nu \ll \kappa$)  \, \,\,\, \\
\mr
  weak driving  & $\langle n\rangle_0\!=\! N\!+\!(N\!+\!1)(\kappa/2\nu)^2 \!+\! \langle n \rangle_{th}$ & $\langle n\rangle_0\!=\![(2N\!+\!1) \!+\!\beta_{th}](\kappa/2\nu) $ \\
  regime  & $W=\eta_c^2 g^2/(2\kappa) (\Omega/\Delta)^2$  &  $W=\eta_c^2 g^2\nu /(2\kappa^2)(\Omega/\Delta)^2$\\
\mr
 saturated   & $\langle n\rangle_0\!=\! N\!+\!(2N\!+\!1)(\kappa/2\nu)^2$ &  $\langle n\rangle_0\!=\!(2N\!+\!1)(\kappa/2\nu)$ \\
  regime & $W=\eta_c^2g^2/\kappa $  &  $W=\eta_c^2g^2\nu/\kappa^2 $ \\
\br
\end{tabular}
\end{indented}
\end{table}
%
%
%
%
 %
%
%
%
%
%
%
%
%
%
Cavity-cooling is most efficient by employing CASC, meaning that the molecule is resonantly excited on the red motional sideband followed by cavity-enhanced decay of rate $\gamma$. In the sideband resolved regime, $\gamma_N=\gamma(2N+1) \ll\nu$, the lower bound for the final occupation number $\langle n\rangle_0$ is then given by $N$, the thermal equilibrium occupation number of the cavity mode, and translates into a final temperature of $T_{\rm f}=(\nu/\omega_{\rm c})T\ll T$. The cooling rate, which scales as $\sim \Omega^2$ for low excitation power is optimized for a value $\Omega_0 \simeq 0.65 \nu$ where cooling rates close to the maximum value of $W\sim \gamma$ can be achieved. Including imperfections like the anharmonicity of the trapping potential characterized in our model by a level dependent shift of the trapping frequency $\delta \nu_n$, the cooling rate decreases as $W_n\sim (\gamma_N/\delta \nu_n)^2$ and eventually re-thermalization processes become dominant.
Our analysis shows that deviations from the ideal case $\langle n\rangle_0\simeq N$ are expected when either the frequency shift exceeds the excitation linewidth $|\delta \nu_n|> \gamma_N$ or when the sideband cooling rate drops below the re-thermalization rate for an insufficient driving strength $\Omega$. Together with the cooling rate also the robustness of CASC is optimized for  $\Omega\sim \Omega_0$.

In the far off-resonant regime $|\Delta|,|\Delta_{\rm c}| \gg \nu,\Omega $ as well as under resonance conditions $\Delta\approx0$ cavity cooling is dominated by the mechanism of $\nabla g$-cooling. In the resonant case, $\nabla g$ cooling is optimized in the fully saturated regime $\Omega \gg \gamma_N,|\Delta|$, where the limit $\langle n\rangle_0\simeq N$ can be reached when the cavity linewidth $2\kappa$ is smaller than the trapping frequency, while the result for Doppler cooling $\langle n \rangle = (2N+1)\kappa/(2\nu)$ applies in the opposite limit. The maximal cooling rate under resonance conditions is $W =\eta_{\rm c}^2 g^2/\kappa$ while in the off-resonant case the molecule is only virtually excited and the cooling rate is reduced by a factor $\Omega^2/\Delta^2$. In this case, to overcome thermally activated heating processes the additional condition $\Omega > \kappa$ is required to reach a final occupation number close to $N$. While in general it might be harder to reach the sideband resolved limit $\kappa \ll \nu$,  $\nabla g $ cooling is by far less sensitive to imperfections as the condition $\delta \nu_n <\gamma_N$ is replaced by $\delta \nu_n <2\kappa$.

In this work we have analyzed cooling rates and final occupation numbers for a fixed set of parameters to obtain a physical understanding of the characteristics and limitations of cavity cooling in different parameter regimes. From this analysis more efficient cooling strategies can be derived by considering the possibilities to switch between different parameter regimes during the cooling process. For example, at the initial stage where the anharmonicity of the trapping potential may be most severe, $\nabla g$ cooling in the Doppler regime offers a very robust cooling method. In a second stage, starting with the precooled molecule localized near the center of the trap, sideband cooling can then efficiently cool the molecule to the ground state. For the design of the optimal cooling strategies we should emphasize that the present setup does not only allow us to change $\Delta(t)$ and $\Omega(t)$ in time, but also offers different possibilities to control $\gamma(t)$, $\Delta_{\rm c}(t)$ and $\kappa(t)$ during the cooling process. For example, using a tunable cavity \cite{WallquistPRB2006,SandbergArXiv2008,PalaciosArXiv2007} the cavity frequency $\omega_{\rm c}(t)$ and therefore also the effective decay rate $\gamma(t)$ can be controlled in a time dependent way. In addition, when the cavity is coupled off-resonantly to another dissipative solid state device, e.g., a Cooper Pair Box, photon states will hybridize with the lossy two-level system and a substantial modulation of the effective photon loss rate can be achieved by  changing the detuning between the two systems.

%
\ack

We acknowledge discussions with D. DeMille, J. Doyle, J. Schmiedmayer, R. Schoelkopf and wish to thank for stimulating input for this work. M. W. acknowledges support from the Marie Curie Research Training Network ConQuest
and EuroSQIP. Work at Harvard was supported by the Packard Foundation and P. R. acknowledges support by the NSF through a grant for the Institute for Theoretical Atomic, Molecular and Optical Physics at Harvard University and Smithsonian Astrophysical Observatory.
%


%
\appendix
%
\section{Molecular master equation in the Lamb-Dicke limit}
\label{app:spectrum}
In this section we present a few details of the Lamb-Dicke expansion (\ref{Eq:rho_m_exp}) of the molecular master equation (\ref{Eq:rho_m})
which was derived in section \ref{sec:Theory}. The first- and second-order terms in the Lamb-Dicke parameters $\eta, \eta_{\rm c}  \ll 1$ read,
\begin{eqnarray}
\fl
\mathcal{L}_1\rho_{\rm m} = -i\frac{\eta\Omega}{2}\Big[\sigma_x(a+a^\dag),\rho_{\rm m} \Big]\nonumber \\
\fl
\qquad + \ \eta_{\rm c} g^2(N+1)\Bigg\{\Big[(\Sigma_-(\nu) a +  \Sigma_-(-\nu) a^\dag)\rho_{\rm m},\sigma_+ \Big]
+ \Big[\Sigma_-\rho_{\rm m}, \sigma_+ (a+a^\dag)\Big] + H.c. \Bigg\}\nonumber \\
\fl
\qquad + \ \eta_{\rm c} g^2 N \Bigg\{\Big[(\Sigma_+(\nu) a +  \Sigma_+(-\nu) a^\dag)\rho_{\rm m}, \sigma_- \Big]
+ \Big[\Sigma_+\rho_{\rm m}, \sigma_-(a+a^\dag)\Big] + H.c. \Bigg\}\,,
\label{Eq:L1}
\end{eqnarray}
and
\begin{eqnarray}
\fl
\mathcal{L}_2\rho_{\rm m} = \eta^2_c g^2(N+1)\Bigg\{\Big[(\Sigma_-(\nu) a +  \Sigma_-(-\nu) a^\dag)\rho_{\rm m},\sigma_+(a+a^\dag)\Big] + H.c. \Bigg\}
\nonumber \\
\fl
\qquad + \ \eta^2_c g^2 N \Bigg\{\Big[(\Sigma_+(\nu) a +  \Sigma_+(-\nu) a^\dag)\rho_{\rm m}, \sigma_-(a+a^\dag)\Big]  + H.c. \Bigg\}\,.
\end{eqnarray}
The operators $\Sigma_\pm (\nu)$ are defined by,
\begin{equation}
\fl
\Sigma_\pm(\nu) \ = \ \int_0^\infty d\tau\, e^{i(\nu \pm(\Delta_{\rm c}-\Delta))\tau}e^{-\kappa \tau} \, \sigma_\pm(-\tau),
\qquad \Sigma_\pm\equiv \Sigma_\pm(0),
\label{Eq:SigmaInt}
\end{equation}
and depend on the Heisenberg operators $\sigma_\pm (t) = e^{iH_{\rm I} t} \sigma_\pm e^{-iH_{\rm I} t}$ with $H_{\rm I} = - (\Delta/2)\sigma_z + (\Omega/2)\sigma_x$ the bare TLS Hamiltonian.
For the evaluation of the integral (\ref{Eq:SigmaInt}) we follow reference~\cite{CiracPRA1995} and write the exponential as
\begin{equation}
\fl
e^{-i H_{\rm I} t } = \cos \left({\bar\Delta t \over 2} \right){\mathbf 1}
+ i \sin \left({\bar\Delta t \over 2} \right) \left[{\Delta \over \bar\Delta} \sigma_z - {\Omega \over \bar\Delta} \sigma_x \right],
\qquad \bar\Delta = \sqrt{\Delta^2 + \Omega^2}.
\end{equation}
Using the algebra of Pauli matrices we arrive at the general form
\begin{equation}
\Sigma_-(\nu) \ = \ B_1(\nu)\sigma_-+B_2(\nu)\sigma_++B_3(\nu)\sigma_z, \qquad \Sigma_+ (\nu)=\Sigma_-^\dagger (-\nu)
\label{Eq:SigmaMinus}
\end{equation}
with the complex coefficients $B_i(\omega)$ given by
\begin{eqnarray}
B_1 (\omega) = {1 \over 4 \bar\Delta^2} \left[{2 \Omega^2 \over \kappa - i (\omega - \Delta_g )}
+ \sum_{s=\pm}{(\Delta + s\bar\Delta)^2 \over \kappa - i (\omega - (\Delta_g + s \bar\Delta))}
\right], \nonumber \\
B_2 (\omega) = {1 \over 4 \bar\Delta^2} \left[{2 \Omega^2 \over \kappa - i (\omega - \Delta_g )}
- \sum_{s=\pm}{\Omega^2 \over \kappa - i (\omega - (\Delta_g + s \bar\Delta))}
\right], \nonumber \\
B_3 (\omega) = {1 \over 4 \bar\Delta^2} \left[{- 2 \Omega \Delta \over \kappa - i (\omega - \Delta_g )}
+ \sum_{s=\pm} {\Omega (\Delta + s \bar\Delta) \over \kappa - i (\omega - (\Delta_g + s \bar \Delta))}
\right],
\label{Eq:Bi}
\end{eqnarray}
with $\Delta_g = \Delta_{\rm c} - \Delta$.
%
\section{Bloch equations}
\label{app:bloch}
For later convenience, we write the Bloch equations (\ref{bloch}) corresponding to the effective TLS dynamics (\ref{Eq:LI})
in the form
\begin{equation}\label{Eq:BlochApp}
\langle \dot{\vec{\sigma}} \rangle \ = \ A \langle \vec\sigma \rangle - \vec\Gamma\,,
\end{equation}
with $\vec \sigma =(\sigma_x,\sigma_y,\sigma_z)^T$ and
\begin{equation}
A  =
\left(
\begin{array}{ccc}
- (\tilde \gamma_N - \gamma_x)/ 2 & \Delta + \tilde \delta - \delta_x & 0 \\
- (\Delta + \tilde \delta + \delta_y) & - (\tilde \gamma_N + \gamma_y)/ 2 & - \Omega \\
\Omega_x  & \Omega + \Omega_y & -\tilde \gamma_N
\end{array}
\right),
\qquad
\vec\Gamma =
\left(
\begin{array}{c}
\Gamma_x \\
\Gamma_y \\
\tilde \gamma
\end{array}
\right).
\label{Eq:AGamma}
\end{equation}
The parameters $\Delta$ and $\Omega$ describe the evolution of a driven free TLS. The interaction of the TLS with the cavity field introduces
energy shifts $\tilde\delta, \delta_x$ and $\delta_y$,
\begin{equation}
\tilde\delta = - g^2 (2N+1) {\rm Im}\{ B_1 (0)\}, \qquad \delta_x = \delta_y = g^2 (2N+1) {\rm Im} \{B_2 (0)\},
\end{equation}
with the complex frequency dependent coefficients $B_i (\omega)$ presented in \ref{app:spectrum}.
Further, the effective decay via the cavity is described by the rates $\tilde\gamma_N = (2N+1)\tilde\gamma, \gamma_x , \gamma_y$,
\begin{equation}
\tilde\gamma = 2 g^2 {\rm Re} \{B_1 (0)\}, \qquad \gamma_x = \gamma_y = 2 g^2 (2N+1) {\rm Re} \{B_2 (0)\} ,
\end{equation}
and $\Gamma_x , \Gamma_y$,
\begin{equation}
\Gamma_x = - 2 g^2 {\rm Re} \{B_3 (0)\}, \qquad \Gamma_y =  2 g^2 {\rm Im} \{B_3 (0)\}.
\end{equation}
Finally there are additional effective Rabi frequencies $\Omega_x , \Omega_y$,
\begin{equation}
\Omega_x = 2 g^2 (2N+1) {\rm Re} \{B_3 (0)\}, \qquad \Omega_y =  2 g^2 (2N+1) {\rm Im} \{B_3 (0)\}.
\end{equation}
From the definition of the $B_i$ (\ref{Eq:Bi}) we find that the non-standard terms $\gamma_{x,y},\delta_{x,y},\Omega_{x,y}$ and $\Gamma_{x,y}$ vanish in the limit $\Omega\ll \kappa$ or else for $\Omega\sim \kappa$ when the additional condition $|\Delta|,|\Delta_{\rm c}|,|\Delta_{\rm c} - \Delta |\gg \Omega$ is fulfilled.

\section{Evaluation of $S_\Omega (\omega)$}
\label{app:S}
In this section we calculate the excitation spectrum defined in equation (\ref{Eq:Somega}),
\begin{equation*}
S_\Omega (\omega) = \frac{\eta^2\Omega^2}{2} \ {\rm Re}  \int_0^\infty d\tau \, {\rm Tr }_{\rm I} \{\sigma_x e^{\mathcal{L}_0\tau} (\sigma_x\rho_{\rm I}^0)  \} e^{i\omega \tau}\,.
\end{equation*}
This expression can be evaluated  using the quantum regression theorem \cite{WallsMilburn}, which relates the two-point correlation function to the
dynamics of the TLS represented by the Bloch equation (\ref{Eq:BlochApp}). After performing the integral we obtain
\begin{equation}
S_\Omega (\omega) =  {\eta^2\Omega^2 \over 2} \ {\rm Re} \left\{  { - h (i \omega)
\over i \omega \ {\rm Det} (i \omega {\mathbf 1} + A) } \right\},
\label{spectrum_S}
\end{equation}
where the function in the numerator is given by,
\begin{equation}
h(i\omega) = (1,0,0) {\rm Ad} (i \omega {\mathbf 1} + A) \left( i
\omega \langle \vec\sigma \sigma_x \rangle_0 + \vec\Gamma \langle
\sigma_x \rangle_0 \right).
\end{equation}
Here we also used the relation $[i \omega {\mathbf 1} + A]^{-1} = {\rm Ad} (i \omega {\mathbf 1} +
A)/{\rm Det} (i \omega {\mathbf 1} + A)$, where Ad(x) is the adjugate matrix of x. By writing the determinant as
\begin{equation}
{\rm Det} (i \omega {\mathbf 1} + A) = (i \omega + \epsilon_0) (i \omega + \epsilon_+) (i
\omega + \epsilon_-),
\end{equation}
with $\epsilon_i$  the eigenvalues of $A$, we clearly see the three-resonance structure familiar from the excitation spectrum of a driven TLS.

Simple analytic expressions for $S_\Omega(\omega)$ can be obtained in certain limits. In the weak driving regime $\Omega \ll \gamma_N$, we can set $\Omega=0$ in the evaluation of $h(i\omega)$ and $\epsilon_i$ and obtain the expression given in equation (\ref{eq:SomegaWeak}). For general $\Omega$ but in the limit $\gamma_N \ll \bar\Delta = \sqrt{\Delta^2 + \Omega^2}$, the spectrum consists of three well separated peaks, with the center frequencies and the widths given by the imaginary and real parts of the $\epsilon_i$. For $\kappa$ such that the Bloch equation (\ref{Eq:BlochApp}) acquires a standard form (see \ref{app:bloch}) we find
\begin{eqnarray*}
 \epsilon_\pm &\simeq&  \pm i \bar\Delta - \bar\gamma/2 ,\qquad
 \bar\gamma = \gamma_N \left( 2 +  \sin^2 \varphi \right)/2 ,\\
\epsilon_0 &\simeq& i 0 - \bar\gamma_0, \qquad \bar\gamma_0 = \gamma_N \left( 1 + \cos^2 \varphi
\right)/2,
\end{eqnarray*}
with $\sin\varphi=\Omega/\bar\Delta$. Using a partial fraction decomposition of equation (\ref{spectrum_S}) and keeping only the lowest relevant orders of $\bar \gamma$ we end up with the simple structure for $S_\Omega(\omega)$ given in equation (\ref{Eq:OmegaSpectrumStrong}).
Since cooling and heating rates are determined by the peaks at non-zero frequencies, $\omega=\pm \bar \Delta$, we can approximately write the spectrum $S_\Omega(\omega)$ as
\begin{equation}
\fl
S_\Omega (\omega \simeq \pm \bar\Delta) =  \frac{\eta^2 \Omega^2}{4}
\frac{\alpha_{\pm} \bar\gamma}{(\omega \mp \bar\Delta)^2+ \bar\gamma^2/4 },
\qquad
\alpha_{\pm} \simeq \cos^2 \varphi \left(\rho_{ee}^0 + {(1 \pm
|\cos\varphi |)^2 \over 2(2N+1) (1 + \cos^2 \varphi)}\right).
\label{Eq:S_SB_MW}
\end{equation}
In section \ref{sec:Discussion} we use this expression to determine cooling and heating rates in the resolved sideband limit for arbitrary values of $\Omega$.
%
%
\section{Evaluate $S_g (\omega)$}
\label{app:g}
Let us now evaluate the $\nabla g$ spectrum (\ref{Eq:Sg}),
using the expression (\ref{Eq:SigmaMinus}) for the generalized TLS operators $\Sigma_\pm (\pm \nu)$.    We write the resulting spectrum on the form,
\begin{equation}
\fl
 S_g (\omega) =
2 \eta_{\rm c}^2 g^2 \ {\rm Re} \Big\{(N+1) \left[ B_1(\omega) \rho_{ee} - B_3 (\omega) \rho_{ge}
 \right]
 + N \left[ B_1(- \omega) \rho_{gg} + B_3 (- \omega) \rho_{eg} \right] \Big\},
\end{equation}
where the complex functions $B_i$ are defined in (\ref{Eq:Bi}), and the TLS steady-state expectation values $\rho_{ij} = \langle i | \rho_{\rm I}^0 | j \rangle $ are derived from the Bloch equations
(\ref{bloch}) under the assumption $\Omega \ll \kappa$ or $\Omega,\kappa \ll |\Delta|,|\Delta_{\rm c}|$. For simplicity we also assume $\gamma_N \ll \bar\Delta$, which is compatible with the constraints for CASC. For transparency we separate the thermal and the driven population,
\begin{eqnarray}
\rho_{ee/gg} = \rho_{ee/gg}^0 \pm \bar\rho_{ee},
\nonumber \\
\rho_{ee}^0 = {N \over 2N + 1}, \qquad \rho_{gg}^0 = {N+1 \over 2N +1}, \qquad  \bar\rho_{ee} = {\Omega^2 \over 2(2N+1)(2 \Delta^2 + \Omega^2)}.
\end{eqnarray}
Regarding the off-diagonal elements $\rho_{eg/ge}$, it can be shown that only the real part contributes to the spectrum in the limit  $\gamma_N \ll \bar\Delta$,
\begin{equation}
{\rm Re }\{ \rho_{eg}\} = {\rm Re} \{\rho_{ge} \} = {2  \Delta \over \Omega } \times \bar\rho_{ee}.
\end{equation}
Without further approximation, we obtain a spectrum with 6 resonances,
\begin{eqnarray}
S_g (\omega) & = &
{\eta_{\rm c}^2 g^2 \over 2 } {\Omega^2 \over \bar\Delta^2 } \left[\
\frac{\kappa (N+1)}{\kappa^2+ (\omega -\Delta_g )^2 } \ + \
 \frac{\kappa N }{\kappa^2+ (\omega + \Delta_g )^2 } \ \right]
\nonumber \\
&+& {\eta_{\rm c}^2 g^2 \over 2 \bar\Delta^2 } \sum_{s=\pm }
 \ \left[ (\Delta + s\bar\Delta)^2 \rho^0_{ee} + \Omega^2 \bar \rho_{ee} \right] \frac{\kappa (N+1)  }
{\kappa^2+ (\omega -(\Delta_g + s \bar\Delta))^2 }
\nonumber \\
& + &
{\eta_{\rm c}^2 g^2 \over 2 \bar\Delta^2 } \sum_{s=\pm } \left[ (\Delta + s\bar\Delta)^2 \rho^0_{gg} - \Omega^2 \bar \rho_{ee} \right] \frac{\kappa N   }
{\kappa^2+ (\omega + (\Delta_g + s \bar\Delta))^2 }
,
\label{Spectrum_D}
\end{eqnarray}
with $\Delta_g \equiv \Delta_{\rm c} - \Delta$.
The two limits of interest are the resolved sideband limit $\kappa \ll \nu$ and the Doppler limit $ \nu \ll \kappa$.


\end{document}